\documentclass{aa} 
\usepackage{graphicx}
\usepackage{txfonts} \usepackage{color} \usepackage{enumitem}
\usepackage{lineno}
\usepackage[dvipsnames]{xcolor}

\usepackage[
  breaklinks = true,
  colorlinks = true,
  urlcolor   = blue,
  citecolor  = blue,
  linkcolor  = blue,
]{hyperref}

\usepackage{bm}

\def \div{{\bm \nabla} \cdot}
\def \curl{{\bm \nabla} \times}
\def \Ktilde {\hat{K}}
\def \xitilde {\hat{\xi}}
\def \xitildejunction {\xitilde_r}

\newcommand{\pot}[2]{{#1\times 10^{#2}}} 
\newcommand{\fracdps}[2]{{{\strut\displaystyle #1}\over{\strut\displaystyle #2}}}
\newcommand{\uvec}[1]{{\hat {\bf #1}}}

\newcommand{\Bvec}{{\vec{B}}}
\newcommand{\Evec}{{\vec{E}}}
\newcommand{\nvec}{{\vec{n}}}
\newcommand{\vvec}{{\vec{v}}}
\newcommand{\jvec}{{\vec{j}}}
\newcommand{\tauvec}{{\vec{\tau}}}
\newcommand{\chiamb}{{\chi_{a}}}
\newcommand{\tauselfsimilarexpansion}{{\tau_{s}}}
\newcommand{\diffflux}{{F_{d}}}

\newcommand{\difffluxnodim}{{D}}
\newcommand{\Bz}{B}
\newcommand{\myBzero}{{{\cal B}}}
\newcommand{\myL}{{\cal L}}

\begin{document}

\title{The Singular Behaviour of Ambipolar Diffusion Revealed by 1D Cartesian Solutions} 

\author{F.~Moreno-Insertis\inst{1,2}
\and
E.~R.~Priest\inst{3}
\and
D.~N\'obrega-Siverio\inst{1,2,4,5}}
\institute{Instituto de Astrof\'\i sica de Canarias, 38205 La Laguna, Tenerife,
  Spain\\ \email{fmi@iac.es}\\ \and Departamento de Astrof\'\i sica, Universidad
  de La Laguna, 38206 La Laguna, Tenerife, Spain. \\  \and School of Mathematics and Statistics, University of St Andrews, St Andrews KY16 9SS, UK\\
\and Rosseland Centre for Solar Physics, University of Oslo, PO Box
  1029 Blindern, N-0315 Oslo, Norway \\ \and Institute of Theoretical Astrophysics, University of Oslo, PO Box 1029 Blindern, N-0315 Oslo, Norway\\
}
\date{February 2026}

\abstract
{Ambipolar diffusion is increasingly recognised as a key process in the lower solar atmosphere. Its highly nonlinear behaviour has many non-intuitive aspects.}
{We seek to (a) study 1D Cartesian ambipolar diffusion near null points;
(b) characterise the nonlinear eigenmodes for ambipolar diffusion; (c)
  propose tests for ambipolar diffusion solvers in MHD codes.}
{(a) Direct analysis is used to find analytical solutions for ambipolar diffusion. 
(b) To study the eigenmodes, we solve the ODE for self-similar solutions of the 1D ambipolar diffusion equation using phase-plane techniques. We also solve the general time-dependent 1D problem for initial conditions of interest. 
(c) We test the Bifrost code by trying to reproduce the behaviour of the eigenmodes.}
%
{(a) A stagnation-point flow solution was found with a uniform flux
  transfer rate across three regions: an external advection region; an
  internal ambipolar diffusion region with magnetic profile $B\propto
  x^{1/3}$; and an innermost Ohmic region with $B \propto x$; in the latter, flux
  annihilation occurs at a rate imposed by the advection. 
(b) Both symmetric and antisymmetric eigenmode solutions to the ambipolar diffusion problem are found with sharp current sheets at the internal nulls. The time evolution of the eigenmodes (pure or perturbed) is probed, showing how higher-order eigenmodes, or perturbed ones, evolve in time towards the lowest-order allowable eigenmodes. 
(c) The Bifrost code reproduces the behaviour of the eigenmodes with excellent accuracy.}
{Stagnation-point configurations exist with ambipolar diffusion
  carrying magnetic flux in an inner layer and serving as an intermediary
  between the external advection and the flux dissipation and
  annihilation at an Ohmic-diffusion core around the null.
Our tests are compatible with the hypothesis
that zero-flux higher harmonics of the
self-similar equation evolve toward either the first symmetric or
antisymmetric harmonic. The self-similar solutions can serve as strong tests for ambipolar diffusion solvers in general MHD codes.
}

\maketitle
\nolinenumbers

\section{Introduction}
\label{sec:introduction}
Ideal magnetohydrodynamics (MHD) is governed by the single-fluid, non-resistive
form of Ohm's law (${\bf E} + \vvec\, \times\, {\bf B} = \bf 0$) and has proved
extremely useful to understand the physics of many simple situations when the
plasma can be assumed to be fully ionised. Yet, for ideal (i.e., non-viscous and non-resistive) MHD, there is no way in
which the magnetic energy, often the largest form of available energy, can be
converted, directly or indirectly, into internal energy via dissipation. Still within the simplest single-fluid approximation
for a fully-ionised plasma but
including the drift between electrons and ions and their mutual friction,
one obtains two additional terms in Ohm's law, namely the Hall term and the
Ohmic diffusion term
\begin{equation}\label{eq:ohm_hall_resistive}
{\bf E} + \vvec\, \times\, {\bf B} + \fracdps{\jvec\times\Bvec}{e\,n_e} =
\fracdps{\jvec}{\sigma}\;,
\end{equation}
\noindent 
where $\sigma = n_e\,e^2/(m_e\,\nu_{ei})$, $\nu_{ei}$ is the electron-ion
collision frequency, and $n_e$ is the number density of electrons
\citep{cowling1957magnetohydrodynamics}. In
Eq.~(\ref{eq:ohm_hall_resistive}), the effect of the electron pressure has
been neglected.  The term on the right of the equation, the Ohmic term, is
due to the electron-ion friction; it leads
to an irreversible increase of the internal energy of the fluid at a rate
$j^2/\sigma$ per unit time. Note that $\vvec$ (in the second term on the left
and in most discussions of Ohm's law in the literature) is the bulk velocity
of the plasma, i.e., in this case, the ion velocity; the Hall term on the
left results from the fact that, if the electron-ion friction were
negligible, the magnetic field lines would be tied to the electron fluid
(given the very small inertia of electrons).

Eq.~(\ref{eq:ohm_hall_resistive}), often neglecting the Hall term,
has been widely used in solving problems for fully-ionised plasmas, like
those in the solar corona.
However, in the lower part of the solar atmosphere, the photosphere and
chromosphere, the plasma is not fully ionised. The neutral atoms can slip
relative to the charged components; yet, there is, also here, a single-fluid
approximation in which the friction between the neutral and charged species
instantaneously adapts to the varying physical situations of the system,
compensating for the difference in the forces acting on the charged and
neutral particles \citep{cowling1957magnetohydrodynamics, braginskii_1965,
  Mitchner_Kruger_1973}. This results in equations for mass conservation and
momentum as in standard MHD for the global fluid
(charged + neutrals), but Ohm's law is changed due to the slippage of the
magnetic field with respect to the neutrals. 

Even if, for simplicity, one ignores the differential effect of the partial
pressures of electrons, ions and neutrals, the ions and neutrals will drift
from each other with a drift speed $\vvec_i - \vvec_n$, which,
  following the assumption of instantaneous adjustment of the friction 
to the driving forces, will be given essentially by equalising the Lorentz
force and the friction between the neutrals and the charged species,
i.e.,
  \begin{eqnarray}\label{eq:drift_speed}
    \vvec_i - \vvec_n &=& \fracdps{\rho_n}{\rho}\,
    \fracdps{1}{n_i\,m_{in}\,\nu^{}_{in}} \;
    \jvec\times\Bvec\;, 
  \end{eqnarray}

\noindent where $m_{in}$ is the reduced ion-neutral mass, $\rho_n$
the mass density of the neutrals, 
$\rho$~the total mass density (ions+neutrals),
$n_i$~the number density of the ions,
$\nu_{in}^{}$~the ion-neutral collision frequency, 
and $\mu_0$ the standard vacuum permeability;
for simplicity, single-charged ions are assumed, so $n_i=n_e$. 
This slippage is called {\it ambipolar diffusion};
correspondingly, the generalised Ohm's law just receives an extra term
compared to Eq.~(\ref{eq:ohm_hall_resistive}) and becomes: 
\begin{equation}\label{eq:ohm_ambipolar}
{\bf E} + \vvec \times {\bf B} + \fracdps{\jvec\times{\bf B}}{e\,n_e} \;=\;
\frac{\jvec}{\sigma} + \mu_0\,\,\chi_a\,B^2\,\jvec_\perp\,,
\end{equation}

\noindent with 
\begin{equation}\label{eq:chia}
\chi_a =
\fracdps{1}{\mu_0}\,\left(\fracdps{\rho_n}{\rho}\right)^2\,\fracdps{1}{n_e\,m_{in}\,\nu_{in}}\;,
\end{equation}

\noindent and the symbol~$\perp$ indicating the component normal to the
magnetic field. In Eq.~(\ref{eq:ohm_ambipolar}), $\vvec$ stands
for the centre-of-mass velocity of the plasma, as in
Eq.~(\ref{eq:ohm_hall_resistive}) but now including both ions
and neutrals.  
The factor $\rho_n/\rho$ can be close to unity in situations
where the degree of ionisation is low, as in the solar photosphere where the ionisation can be 
as low as $10^{-4}$. The plasma becomes completely ionised only above the upper chromosphere.

Taking the curl of Eq.~(\ref{eq:ohm_ambipolar}) and using Faraday's law
turns it into a generalised induction equation which, 
for the simple case in which the Hall term can be neglected,
can be written as follows: 
\begin{equation}\label{eq:induction_ambipolar_3D}
\frac{\partial {\bf B}}{\partial t}=
\curl \left[\strut \vvec \times {\bf B}\,-\, \eta\,\curl \Bvec \, -
\,\chi_a\,B^2\,(\curl \Bvec)_\perp \right]\;,
\end{equation}
where  $\eta = 1/(\sigma\,\mu_0)$. 
The resulting internal energy equation in the presence of both Ohmic and ambipolar diffusion, but disregarding viscous dissipation, has the form
\begin{equation}\label{eq:internal_energy}
\rho\,\frac{D\epsilon}{Dt}\;=\; - p\,\div\vvec \;+\; \frac{j^2}{\sigma}\;+\;
\mu_0\,\chi_a\, B^2\,{j_\perp\hskip -2pt}^2\;,
\end{equation}
with $\epsilon$ the internal energy per unit mass
and the symbol $D/Dt$ indicating the material derivative
$(\hskip 1pt{\partial} / {\partial} t + \vvec \cdot \nabla\hskip 1pt)$.  Thus, both Ohmic
and ambipolar diffusion can convert magnetic energy into heat as the magnetic
field evolves, the former because of the electron-ion friction, the
  latter because of the friction between neutrals and charged species.

The magnitude of the ambipolar resistivity $\chi_a\,B^2$
in the solar atmosphere is debated; under the strong and often unrealistic
assumption of LTE for the ionisation level of Hydrogen, one would conclude
that, in the upper solar photosphere, ambipolar diffusion can be a factor of
$10$ larger than normal Ohmic diffusion; in the chromosphere the factor
becomes $10^3$ in the quiet Sun or $10^5$ in an active region
\citep{khomenko12a,priest14a}; yet, non-equilibrium ionisation effects can
drastically change the importance of this effect (see
\citealt{Nobrega-Siverio_etal_2020a, Wargnier_etal_2022}; see also
\citealt{Hillier_2024}.)

Ambipolar diffusion differs in a number of important ways from normal Ohmic
diffusion: 

(i) it is a nonlinear process in $B$, and so has many counter-intuitive effects;

(ii) it vanishes at null points unless the current becomes singular;
in fact, it tends to steepen field profiles that contain a null into
  sharp current sheets (see also Sec.~\ref{sec:general_solutions} and
\ref{sec:discussion}, where the possibilities of having annihilation and
reconnection at the null via a small Ohmic region within an ambipolar-dominated domain are discussed.)

(iii) the time-scales are very different;  for a length-scale $L_0$ and magnetic field $B_0$ the Ohmic diffusion time-scale is
\begin{equation}\label{eq:timescale_Ohmic}
\tau_d=\frac{L_0^2}{\eta}\;,
\end{equation} 
whereas the ambipolar diffusion time-scale is
\begin{equation}\label{eq:timescale_ambipolar}
\tau_a=\frac{L_0^2}{\chi_a\, B_0^2} \;.
\end{equation}
Thus, ambipolar diffusion is fastest in regions of high magnetic field.  When
ambipolar diffusion is stronger than Ohmic diffusion (as assumed in this
paper), the former will be the dominant diffusion process in the induction
equation, although Ohmic diffusion can still be relevant around null points.

In this paper we concentrate on a simple {\it plane-parallel} situation in
which ${\bf B} = B(x,t) \,\uvec{y}$ and the coefficients $\eta$ and
$\chi_a$ are uniform; we shall also be assuming a static plasma 
(except in Sec~\ref{sec:stagnation-point}), so that the induction
equation~(\ref{eq:induction_ambipolar_3D}) reduces to
\begin{equation}\label{indeq3}
\frac{\partial B}{\partial t}\,=\,\eta\,\fracdps{\partial^2 B}{\partial x^2}
\,+\, 
\chi_a\, \frac{\partial }{\partial x} \left(B^2 \frac{\partial B}{\partial x}
\right),  
\end{equation}
which is a nonlinear diffusion equation. 
We aim to determine explicit analytical
solutions and numerical examples for this equation in order to
highlight the mathematical and physical consequences of ambipolar
diffusion in a one-dimensional Cartesian geometry and to explore
  their possible use in benchmarking ambipolar diffusion solvers in MHD
  codes. 

This paper builds on work presented in a previous publication
\citep{Moreno-Insertis_etal_2022}, where we studied solutions to the
equivalent problem in a cylindrically symmetric situation and devised tests
for MHD numerical codes that include an ambipolar diffusion solver. The
ambipolar diffusion problem in either geometry admits signed self-similar
solutions which are counterparts to the basic family of unsigned solutions
for a large class of nonlinear 
diffusion problems discovered by \citet{Zeldovich_Kompaneets_1950,
  Barenblatt1952}; see also \citet{Pattle1959, Zeldovich_book1967,
  Vazquez_2007}.  The
mathematical properties of all these solutions have been studied in depth in
the applied mathematics literature (see, e.g., \citealt{Grundy1979,
  Hulshof_1991, Vazquez_2007}; and the brief summary in
  \citealt{Moreno-Insertis_etal_2022}).

The study of ambipolar diffusion in plasma physics started long ago \citep{cowling1957magnetohydrodynamics} as did its applications to astrophysical plasmas \citep[e.g.,][]{Parker_1963}. Such applications 
are numerous and cover very different cosmic contexts. 
In the solar photosphere and chromosphere, the importance of ambipolar
diffusion was stressed by \citet{priest82} in his monograph on MHD; 
the interest on this phenomenon has gradually increased until the present. 
Solar physics applications include the excitation, propagation
and mode-conversion of MHD waves
\citep[e.g.,][]{kumar03,vranjes08,soler10,zaqarashvili11,cally18},
Kelvin-Helmholtz instability \citep{soler12,hillier19,hillier24}, Rayleigh-Taylor instability
\citep{ruderman18}, heating of the chromosphere \citep{Goodman_2000,
  Goodman_2004, khomenko12a}, 
magnetic reconnection \citep{Parker_1963,zweibel89, Heitsch_Zweibel_2003a, 
sakai09,zweibel09,leake12a,leake13a,Snow_etal:2018} 
and the tearing-mode instability \citep{Ni_etal_2015,Ni_etal_2016,Ni_etal:2022,Cheng_etal:2024}. 
Detailed computational modelling of its effects has been presented for 
the emergence of magnetic flux from below the solar photosphere
\citep{leake06,arber07,Arber_etal_2009, leake13b, 
Nobrega-Siverio_etal_2020b}, for the fine structure of
the solar photosphere and chromosphere
\citep{martinez12a, martinez17a}, particularly regarding spicules 
\citep{depontieu17a,martinez17b,martinez18a,Martinez-Sykora_etal:2020}, formation of structure in solar
prominences  \citep{khomenko14a, khomenko_etal_2016},
and magnetoconvection \citep{khomenko18a}. 

In non-solar astrophysics, ambipolar diffusion is important in many similar
physical processes, but under different parameter regimes, such as the
structure of interstellar shock waves \citep{draine86,draine93} and the
formation of current sheets in the interstellar medium
\citep{Zweibel_Brandenburg_1997}.  It has also been considered as a mechanism for
diffusion of magnetic flux in dense molecular clouds
\citep{mestel56,spitzer62}, where it plays a key role in star formation
\citep{shu83,crutcher_99, MacLow_Klessen_2004, Nakamura_Li_2008,
  ballester18}, magnetic braking
\citep{mouschovias_paleologou_1979, li_etal_2014}, and the
  formation of filamentary structure \citep{hennebelle13}.  
Further numerical models in 2D and 3D that study the 
    effects of ambipolar diffusion in molecular clouds are those of
    \cite{kunz_mouschovias_2010} and \cite{tritsis_2025}.
In planetary ionospheres,
ion-neutral coupling and therefore ambipolar diffusion is important in
creating global currents and neutral winds \citep[e.g.,][]{bauer73,
  pfaff12,ballester18}.

In this paper, we start by giving an overview of past results
  concerning the $B\sim x^{1/3}$ profiles that naturally arise around null
  points in ambipolar diffusion contexts  (Sec.~\ref{sec:nulls}) and then
(Sec.~\ref{sec:stagnation-point}) introduce a generalisation of the
stagnation-point flow that includes both ambipolar and Ohmic diffusion terms.
A second part of the paper is then devoted to calculating
  self-similar solutions with compact support, both symmetric and
  antisymmetric, which have been shown in the literature to constitute a
  countable set (Sec.~\ref{sec:self-similar}). Numerical solutions starting
from various initial conditions permit an initial exploration of the
  stability of the self-similar solutions
(Sec.~\ref{sec:solutions_overview}). Sec.~\ref{sec:bifrost}
  illustrates the usefulness of the solutions in testing numerical codes
with ambipolar diffusion solvers. Finally,
  Secs.~\ref{sec:discussion} and \ref{sec:conclusions} contain a
  discussion and 
  conclusions, including multi-fluid considerations on the 
  important differences between ohmic and ambipolar diffusion regarding
  magnetic flux conservation and annihilation.  

\section{General results from conservation theorems}\label{sec:general_solutions}

Consider the induction equation for a static plasma with Ohmic and
ambipolar terms, i.e., Eq.~(\ref{eq:induction_ambipolar_3D}) with
  $\vvec=0$. 
Integrating it over a surface S bounded by a curve C which does not change in
time and using Stokes' theorem gives
\begin{equation}\label{eq:flux_integral}
\frac{d}{dt}\int_S {\bf B} \cdot \nvec\, dS = -\,\mu_0\int_C
  \left(\eta\,\jvec + \chi_a\, B^2\, \jvec_\perp\right)
  \cdot \tauvec \, dl  \,,
\end{equation}
with $\nvec$ the unit normal vector to the surface and $\tauvec$ the unit
tangent vector to the curve. 
For the case of a one-dimensional field, \hbox{$\Bvec = B(x,t)\,\uvec{y}$,}
still with $\vvec=0$, the induction equation becomes 
Eq.~(\ref{indeq3}), which 
 may be rewritten
\begin{equation}\label{eq:ad_diffusion_equation}
\frac{\partial \Bz}{\partial t} =-\frac{\partial \diffflux}{\partial x}
\end{equation}
in terms of the total diffusive flux (i.e., 
the magnetic flux transfer rate in the diffusive regime), namely,
\begin{equation}\label{eq:diff_flux_definition}
\diffflux(x,t) = -\eta\, \frac{\partial \Bz}{\partial x} - \chi_a\,
\Bz^2\frac{\partial \Bz}{\partial x}\;, 
\end{equation}
which is the sum of the Ohmic 
and ambipolar 
diffusive fluxes. Eq.~(\ref{eq:ad_diffusion_equation}) may be integrated over $x$ from $a$ to
$b$, say, to give the time-variation of the magnetic flux $\Phi$ as
\begin{equation}\label{eq14b}
\frac{d\Phi}{dt}
= 
\frac{d}{dt}\int_a^b \Bz \ dx= -[\diffflux]_a^b\;, 
\end{equation}
so that naturally the rate of change of magnetic flux is the difference between the total diffusive fluxes at the two boundaries.

\begin{figure}
{\centering
\includegraphics[width=0.35\textwidth]{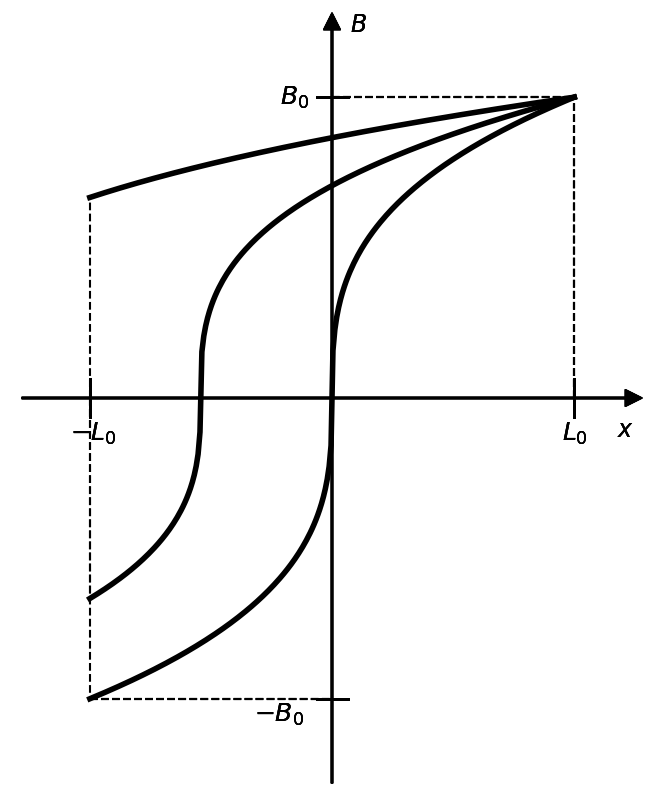}
\caption{A sketch of the solution for steady-state ambipolar diffusion joining $B=B_1$ at $x=-L_0$ to $B=B_0$ at $x=L_0$ for three different values of $B_1$, namely, $2 B_0/3$ (top curve), $-2 B_0/3$ (middle curve), and $-B_0$ (bottom curve).
\label{fig:fig1}}}
\end{figure}

From Eq.~(\ref{eq:ad_diffusion_equation}), the steady-state
  solution for the purely ambipolar problem requires 
$F_d$ to be uniform,
    i.e., following Eq.~(\ref{eq:diff_flux_definition}), that $d \Bz^3/dx$ be
    uniform. When joining two values ($B_0$ and $B_1$, say) of the same
sign at $x=\pm L_0$, the magnetic field is well behaved (as in the top curve
of Fig.~\ref{fig:fig1}). When $B_0$ and $B_1$ have opposite sign, an
$x^{1/3}$ singularity appears where the solution crosses the $x$-axis (as in
the two lower curves of Fig.~\ref{fig:fig1}). In the next subsection, we
focus on the antisymmetric case ($B_1 = -B_0$, namely, the lowest curve in
Fig.~\ref{fig:fig1}), while in Sec.~\ref{sec:stagnation-point} a stagnation
flow solution is introduced, all within the single-fluid approach of the
generalised Ohm's law; a discussion of some multi-fluid considerations
underlying the MHD solution and possibly limiting it is given in
Sect.~\ref{sec:discussion}.

\subsection{Diffusive behaviour near null points, flux transfer and
  annihilation}\label{sec:nulls} 

\begin{figure}
{\centering
\includegraphics[height=7cm]{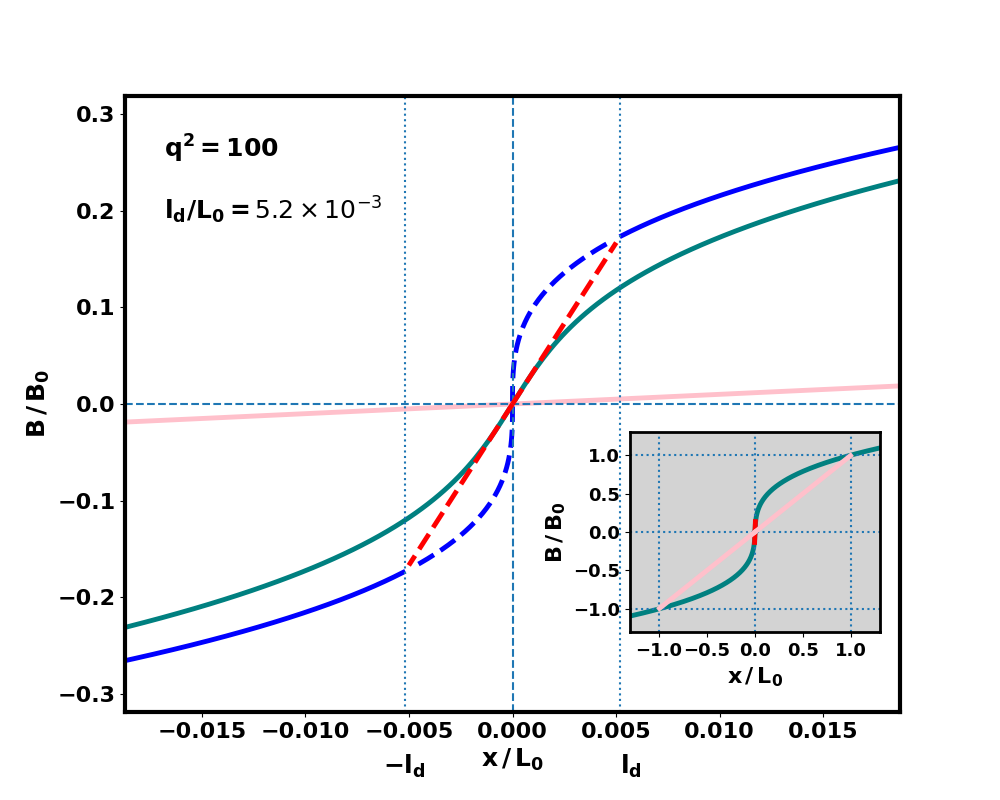}
\caption{The solution to the mixed Ohmic and ambipolar diffusion problem of Eqs.~(\ref{eq:ad_diffusion_equation})-(\ref{eq:diff_flux_definition}), in which the inset shows the whole range $-L_0 \leq x\leq L_0)$ and the main frame shows just the central domain $-0.018L_0 \leq x\leq 0.018L_0)$]. Green: the exact solution (\ref{eq:stationary_Parker_2}); blue: the purely ambipolar solution (\ref{eq:Parker_x_to_onethird}); dashed red: the linear, purely Ohmic solution with slope at the origin equal to that of the full solution (\ref{eq:Parker_linear}); pink: the linear purely Ohmic solution linking the points $(-L_0, -B_0)$ and $(L_0, B_0)$. The dotted vertical lines mark the locations $x=\pm l_d$. The blue and green curves and the red segment appear superimposed in the inset.}
\label{fig:fig2}}
\end{figure}

\subsubsection{\citeauthor{Parker_1963}'s 
\citeyear{Parker_1963} solution}\label{sec:the_Parker_solution} 

Null points play a special role in this problem: the ambipolar diffusive flux
$-\chi_a\,B^2\,\partial B/\partial x$ vanishes at them so that only the Ohmic
diffusion term (or perhaps some collisionless diffusion process) can lead to
magnetic flux annihilation.  This was already recognised in a seminal paper
by \citet{Parker_1963}, who considered the problem of stationary solutions to
the diffusion equation
(\ref{eq:ad_diffusion_equation})-(\ref{eq:diff_flux_definition}). Assuming
uniform coefficients $\chi_a$ and $\eta$, he obtained an exact solution of
the general form
\begin{equation}\label{eq:stationary_Parker}
c_2\, x={\textstyle{\frac{1}{3}}}\chi_a \,\Bz^3+\eta\, \Bz,
\end{equation}
with $c_2$ an arbitrary integration constant,
in which the null is assumed to be located at $x=0$. Clearly, that equation
gives a behaviour of the kind $\Bz \propto x^{1/3}$ wherever ambipolar
diffusion dominates Ohmic diffusion, and a linear behaviour, $\Bz \propto x$,
in the opposite case. Assume that the case at hand is ambipolar-diffusion
dominated almost everywhere, i.e., more precisely, assume that, for a generic
point at a distance $L_0$ from the origin ($x=L_0$) with field strength $B_0$, the ratio of the first to the second
term on the right hand side of (\ref{eq:stationary_Parker}) is large:
\begin{equation}\label{eq:q2}
q^2 = \fracdps{\chiamb\,B_0^2}{\eta} = \fracdps{\tau_d}{\tau_a} \gg 1\;,
\end{equation}

\noindent with $\tau_d$ and $\tau_a$ as defined in
Eqs.~(\ref{eq:timescale_Ohmic}) and (\ref{eq:timescale_ambipolar}).  
In that case, (\ref{eq:stationary_Parker}) can be turned into
\begin{equation}\label{eq:stationary_Parker_2}
\fracdps{x}{L_0} = \fracdps{\Bz}{B_0} \;\left(\fracdps{3}{q^2} + \fracdps{\Bz^2}{B_0^2}\right)  \left(\fracdps{3}{q^2} + 1 \right)^{-1}\; \approx\; \fracdps{\Bz}{B_0} \;\left(\fracdps{3}{q^2} + \fracdps{\Bz^2}{B_0^2}\right)
\end{equation}

\noindent (the green curve in Fig.~\ref{fig:fig2}, both in the main frame and in the inset). \cite{Parker_1963} noted that the solution 
\begin{equation}\label{eq:Parker_x_to_onethird}
\fracdps{\Bz}{B_0} = \left(\fracdps{x}{L_0}\right)^{1/3}
\end{equation}
(the blue curve in the main frame of the figure) is valid everywhere except in the immediate neighbourhood of the null point, which is Ohmic-diffusion dominated: in the narrow range 
$-l_d \le x \lesssim l_d$, with 
\begin{equation}\label{eq:diffusive_length}
l_d = L_0\,\left(\frac{3}{q^2}\right)^{3/2}\, ,
\end{equation}
the solution is instead linear:
\begin{equation}\label{eq:Parker_linear}
\fracdps{\Bz}{B_0} = \fracdps{x}{L_0}\, \fracdps{q^2}{3}\;
\end{equation}

\noindent (the red dashed straight line in the main frame of Fig.~\ref{fig:fig2}
and the red segment in the inset). Parker's conclusion was that ambipolar
diffusion is not the dominant mechanism near the null, and that flux transfer
and annihilation across the null are possible by Ohmic diffusion.
Yet, due to the presence of ambipolar diffusion in most of the domain, the slope of the $B$-profile at the null, $(B_0/L_0)\, q^2/3$ (the red dashed line in the figure),  is much larger than given by a simple Ohmic diffusion solution in the whole domain (having slope $B_0/L_0$, the pink line in the main frame and in the inset). Correspondingly, Parker calculated that the total Joule dissipation in the small range $-l_d \le x \le l_d$ for the Ohmic-dominated part of the whole solution is a factor $q$ larger than that of the simple pure Ohmic solution in the range $(0, L_0)$. 

\subsubsection{The flux transfer rate in the $B\propto x^{1/3}$
    profile}\label{sec:the_singular_profile}

As an extension of the foregoing, we note the remarkable fact that in an
ambipolar-diffusion dominated region with profile $\Bz \propto x^{1/3}$
(whether stationary or otherwise) the flux transfer rate (or diffusive flux)
$\diffflux$ is uniform: in that region the electric current has a steep
profile $j_y \sim x^{-2/3}$ approaching a singularity as $x\to 0$, but the
field goes to zero, so $\diffflux = -(\chi_a /\mu_0) j_y \Bz^2 = $
constant. Therefore, the ambipolar term will be passing flux to the
Ohmic-dominated region at a uniform rate thanks to the steepness of the
profiles in spite of the smallness of $\Bz$. Flux dissipation and
reconnection are then possible in the region around the null at rates
determined by the ambipolar diffusion, even if not directly in the
ambipolar-diffusion dominated part of the domain.

In the stationary case, we can show that the flux transfer across the
  null in the combined Ohmic+ambipolar problem has the same intensity as for
  a purely ambipolar diffusion solution (\ref{eq:Parker_x_to_onethird}); to
  that end, we compare the Ohmic flux transfer rate
$\,\eta\, \partial \Bz/\partial x$ for the actual solution near the centre
(\ref{eq:Parker_linear}), namely, $\eta\,B_0\, q^2 /(3\, L_0)$, with the
transfer rate that would apply if the singular, purely ambipolar diffusion
solution were valid there, namely $\chi_a\,\Bz^2\,\partial \Bz/\partial x$
with $\Bz$ 
given by Eq.~(\ref{eq:Parker_x_to_onethird}): the two values are identical as
can be easily checked by using the definition of $q^2$. So, even when Ohmic
diffusion prevails near the null, the flux transfer rate (and possible
annihilation properties) is determined by ambipolar diffusion.

\subsection{Stagnation-point flow}
\label{sec:stagnation-point}

\begin{figure}
{\centering
\includegraphics[width=0.40\textwidth]{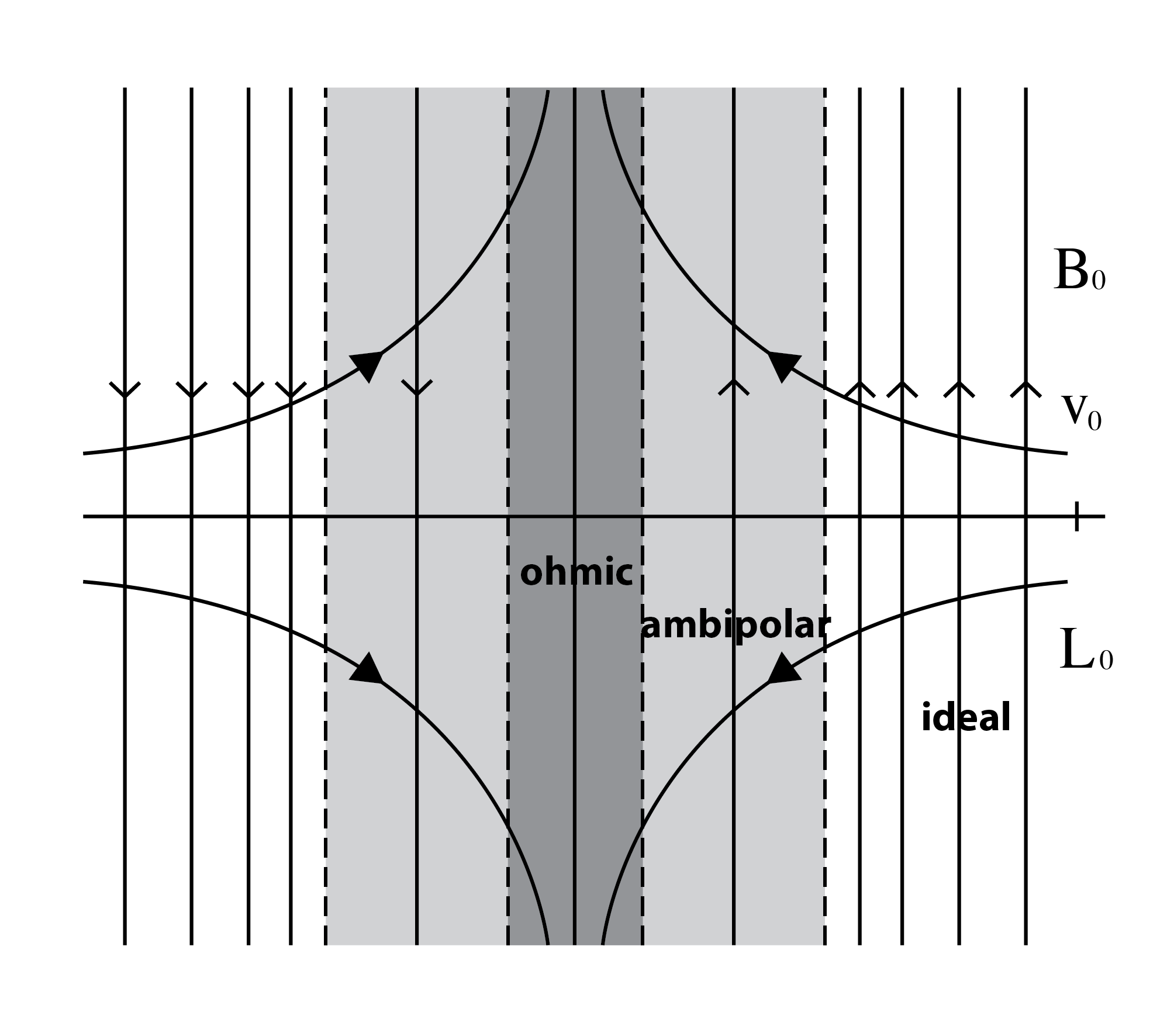}
\includegraphics[width=0.45\textwidth]{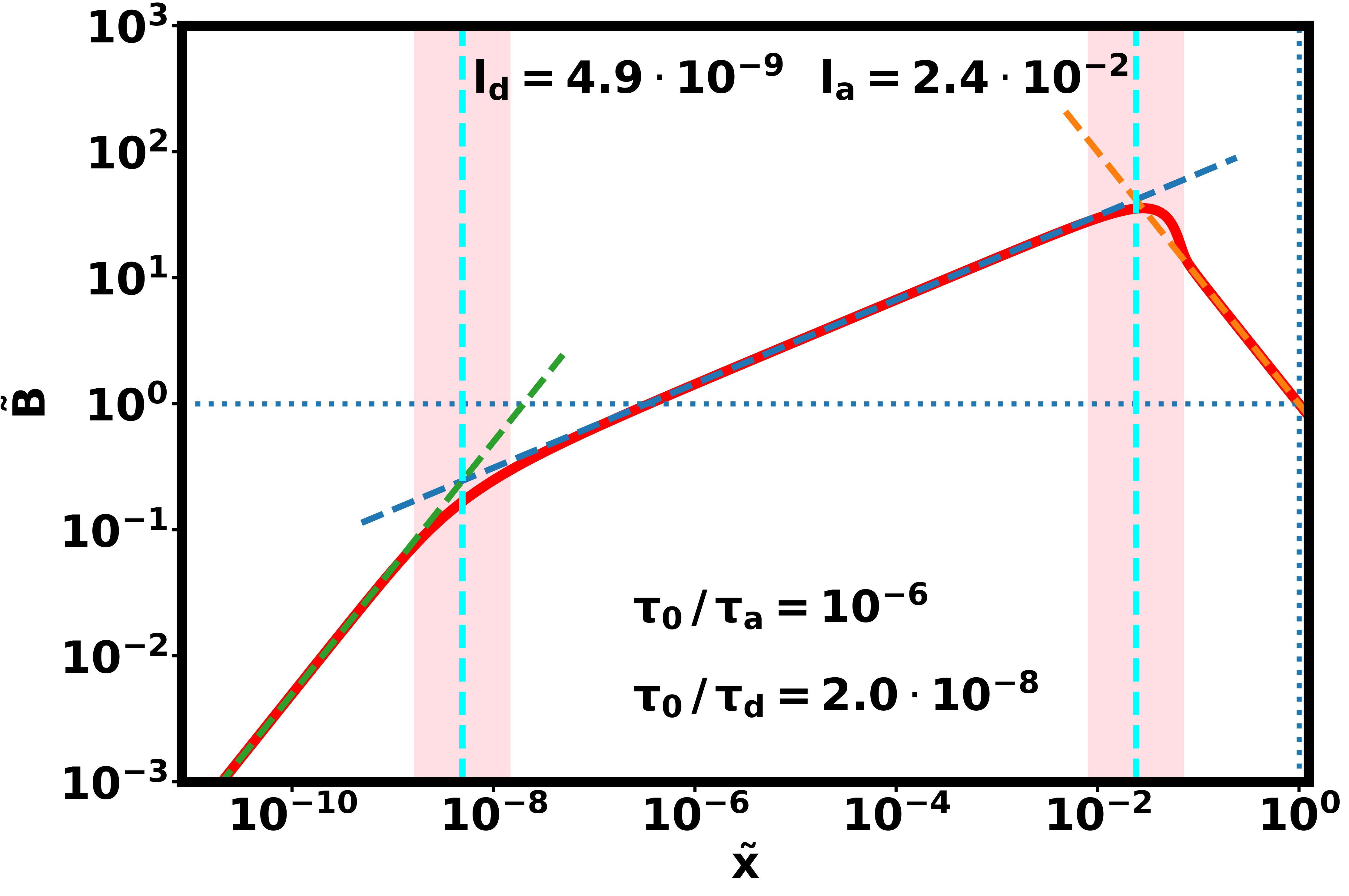}
\includegraphics[width=0.42\textwidth]{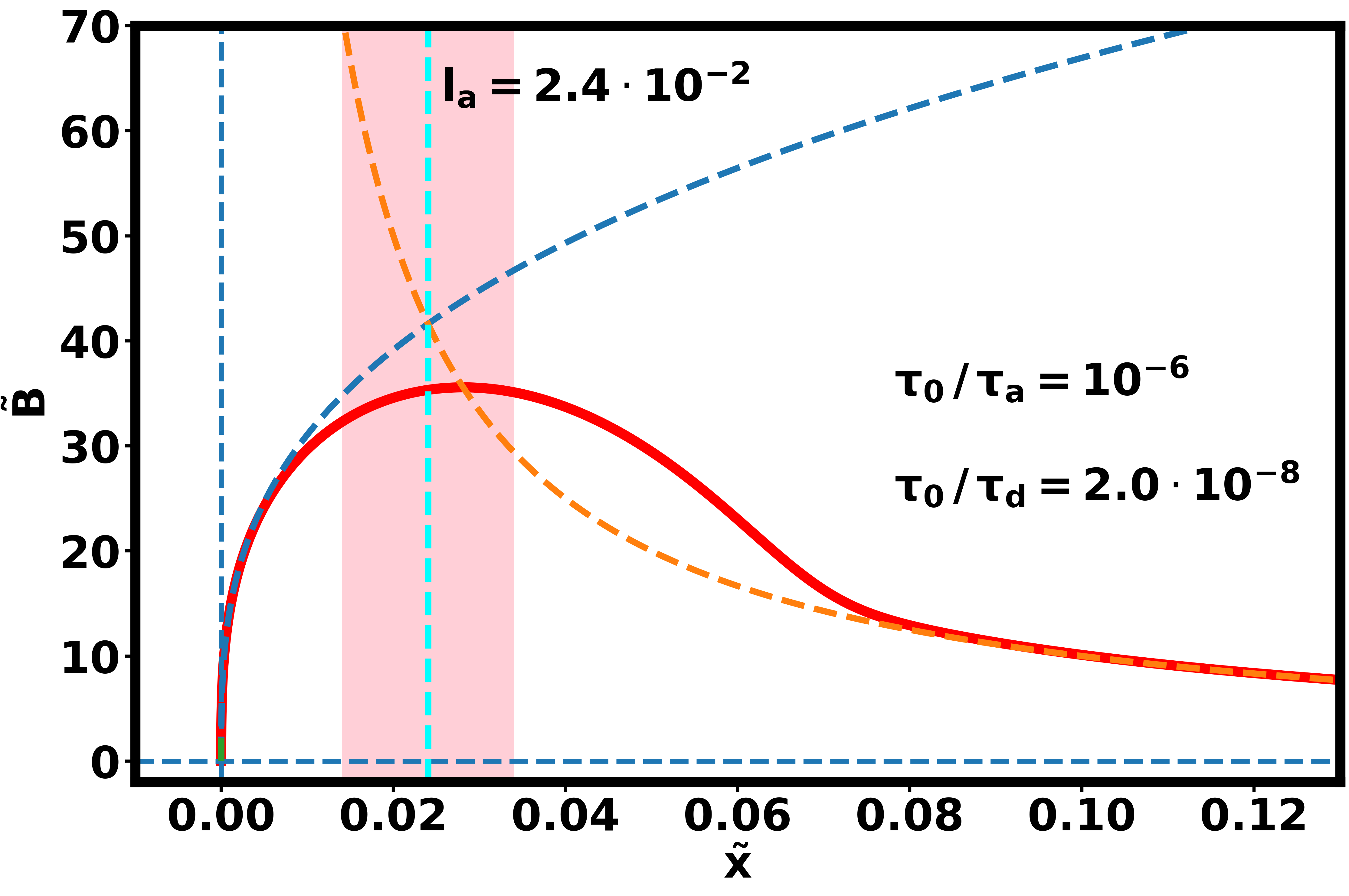}
\caption{Upper panel: a schematic of a stagnation-point flow configuration,
  showing magnetic field lines (light-headed arrows) for a one-dimensional
  current sheet together with a stagnation-point flow (solid-headed
  arrows). The ideal, ambipolar and Ohmic regions are indicated. Middle and
  lower panels: solution of the dimensionless stagnation flow equation
  (Eq.~\ref{eq:stagnation_dimensionless}) with logarithmic (middle panel) and
  linear axes (lower panel). Red solid: solution for $\tilde{B}(\tilde
  x)$. Dashed: the asymptotic regimes for the purely advection-dominated
  (orange), ambipolar-dominated (blue) and Ohmic-dominated (green) regimes,
  behaving like $x$, $x^{1/3}$ and $x^{-1}$, respectively.
  The cyan vertical dashed lines mark $l_a$ and $l_d$, with a pink vertical
  band around them to highlight the transition between different asymptotic
  regimes ($l_d$ and the asymptotic Ohmic curve are only shown in the log
  plot).  
\label{fig:stagnation}}}
\end{figure}

\cite{parker73} considered a kinematic solution for the effect of two-dimensional stagnation-point flow on the induction equation alone including Ohmic diffusion, whereas \cite{sonnerup75} extended this to a three-dimensional stagnation-point flow and they showed that it represents one of the few exact solutions of MHD since it also satisfies the equation of motion.  We can now extend the  2D solution to include ambipolar diffusion as follows.

Consider unidirectional oppositely-directed magnetic fields,  $\Bz(x)\,\uvec{y}$, that are driven together by a stagnation-point flow of the form: 
\begin{equation}\label{eq:aux1}
v_x = -v_0\frac{x}{L_0}, \ \ \ \ \ v_y=v_0\frac{y}{L_0},
\end{equation}
with $v_0 >0$, as indicated in the upper panel of Fig.~\ref{fig:stagnation}.
In a steady-state situation, Ohm's law including advection, ambipolar diffusion and Ohmic diffusion may be written
\begin{equation}\label{eq:Ohm}
E-\frac{v_0x}{L_0}\Bz=\eta \frac{d\Bz}{dx} +\chi_a \Bz^2\frac{d\Bz}{dx};
\end{equation}
given the geometry and the condition $\curl \Evec = \bf 0$, the electric
field must be uniform. Also, we will seek antisymmetric solutions for
$\Bz(x)$ and will thus discuss the solution for $x>0$ only. To avoid
  infinities for large $x$,  
the effects of ambipolar and Ohmic diffusion must be negligible at large $x$. The solution in the asymptotic region (the {\it ideal region}) is just given by
\begin{equation}\label{ideal}
\Bz=\frac{E\, L_0}{v_0 \,x}=B_0\frac{L_0}{x},
\end{equation}
\noindent where $L_0$ is the coordinate of any point in that region and 
$B_0$ the value of the field at it if exactly on the asymptotic
  curve. The natural time-scale for the ideal region is the advective
time-scale defined by  
\begin{equation}\label{eq:advective_time_scale}
\tau_0 = 
\frac{L_0}{v_0}.
\end{equation}

\noindent Suppose that the magnetic field vanishes at $x=0$; 
as one goes
inward from 
$x=L_0$ toward $x=0$, at some point the ambipolar and Ohmic diffusion
effects will become relevant and eventually dominant. If, as in the
foregoing section, the Ohmic diffusion overcomes the effects of ambipolar
diffusion in a narrow region close to the centre only, then we will have
three distinct regions in the problem, the outer one (the ideal solution), a
middle one dominated by ambipolar diffusion and the central one, around $x=0$,
dominated by Ohmic diffusion. Assume the transition between the ideal and
ambipolar-diffusion dominated regions occurs around $x=\,$O$(l_a)$, whereas
the 
transition between ambipolar and Ohmic diffusion dominated regions is located 
at $x=\,$O$(l_d)$. 

Clearly, wherever ambipolar diffusion dominates, the
solution is given by:

\begin{equation}\label{ambipolar}
\frac{\Bz}{B_0}=\left(\frac{3Ex}{\chi_aB_0^3}\right)^{1/3}=\left(3\,\frac{\tau_a}{\tau_0} \frac{x}{L_0}\right)^{1/3},
\end{equation}

\noindent so, the extent ($l_a$) of the region dominated by ambipolar diffusion
 is given by the value of $x$ that equates the ideal and ambipolar
fields (\ref{ideal}) and (\ref{ambipolar}), namely, 
\begin{equation}\label{eq:length_ratio}
\frac{l_a}{L_0} 
=
\left(\frac{B_0^2\,\chi_a}{3\,v_0\,L_0}\right)^{1/4}=\left(\frac{1}{3} \frac{\tau_0}{\tau_a}\right)^{1/4}.
\end{equation}

\noindent with $\tau_a$ as defined in Eq.~(\ref{eq:timescale_ambipolar}),
but now with the symbols $L_0$ and $B_0$ as used in the present
  section (see Eq.~\ref{ideal}).
The condition that $L_0$ is in the ideal region, i.e., $l_a \ll L_0$, is
equivalent to the condition 
$(\tau_0 /\tau_a)^{1/4} \ll 1$. 

In the Ohmic diffusion region, on the other hand,
\begin{equation}\label{ohmreg}
\frac{\Bz}{B_0}=\frac{E\,x}{B_0\,\eta}=\frac{\tau_d}{\tau_0}\,\frac{x}{L_0}, 
\end{equation}
with $\tau_d$ as defined in Eq.~(\ref{eq:timescale_Ohmic}), but now
  with the symbol $L_0$ defined by
  Eq.~(\ref{ideal});
so the location ($l_d$) where the transition between the ambipolar
and Ohmic dominated regions occurs  
is given by equating the fields in the ambipolar and Ohmic diffusion regions (\ref{ambipolar}) and (\ref{ohmreg}), namely,
\begin{equation}\label{eq24}
\frac{l_d}{L_0}
=
\left(\frac{3\,\eta^3}{L_0^2\,E^2\,\chi_a}\right)^{1/2} =
\frac{1}{3}\,\left(\frac{3\,\tau_a}{\tau_d}\right)^{3/2}\,\frac{\tau_0}{\tau_a}\;. 
\end{equation}

\noindent From Eqs.~(\ref{eq:length_ratio})~and~(\ref{eq24}), whenever the
condition $l_a \ll L_0$ is  
satisfied, then the condition $l_d \ll l_a$ will also be satisfied in as far
as the weaker additional requirement is fulfilled: 
\begin{equation}\label{eq:AD_to_Ohm_length_ratio}
\left(\fracdps{\tau_a}{\tau_d}\right)^{3/2}\;
\left(\fracdps{\tau_0}{\tau_a}\right)^{3/4}\;\ll 1\;.
\end{equation}

Of particular interest is the fact that, according to Ohm's law
  (Eq.~\ref{eq:Ohm}),  the flux transfer rate (i.e., the sum of the two
diffusive terms and the 
advection term) is uniform in the whole domain. On the other hand, in each of
the three regions just discussed only one of those terms is significant, and
it therefore carries the whole magnetic flux in that region. Finally, in this
problem, the magnetic flux inflow is determined by the external region, i.e.,
by advection.  We see, therefore, that the advection inflow sets the
value of the electric field and, with it, the overall flux transfer rate; the
flux is then transferred by ambipolar and Ohmic diffusion in their respective
domains at that same rate, which is also the flux annihilation rate in the
innermost region.

To obtain the full magnetic field profile one can use a 
dimensionless version of Eq.~(\ref{eq:Ohm}), namely:
\newcommand{\Btz}{{\tilde{B}}}
\begin{equation}\label{eq:stagnation_dimensionless}
1 - \tilde{x}\,\Btz = \left(\frac{\tau_0}{\tau_d} \; + \;  
\frac{\tau_0}{\tau_a}\,
\Btz^2
 \right)
\frac{d\Btz}{d\tilde{x}} \;,
\end{equation}
with $\tilde{x} = x/L_0\,$ and $\Btz = \Bz\,v_0/E$, and solve it
  using simple numerical means. The profile is shown in
Fig.\ref{fig:stagnation} (middle and lower panels) for given choices of the
parameters $\tau_0/\tau_d$ and $\tau_0/\tau_a$.  In this solution, the
magnetic field is advected in by the flow from a distance $x=L_0$ to around
$l_a$, and then it diffuses inwards due to the diffusive flux
(Eq.~\ref{eq:diff_flux_definition}). First of all, this field is carried by
the ambipolar diffusive flux between about $l_a$ and $l_d$, and secondly by
the Ohmic diffusive flux from about $l_d$ to the origin, where it is
annihilated as it cancels with equal and opposite flux from $x<0$.

\section{General self-similar solutions}
\label{sec:self-similar}

The analysis of self-similar solutions to the purely ambipolar diffusion
equation, Eqs.~(\ref{eq:ad_diffusion_equation})-(\ref{eq:diff_flux_definition})
  with $\eta=0$, follows the same general scheme that we developed for
symmetric solutions in a cylindrical geometry
\citep{Moreno-Insertis_etal_2022}. However, here we will also be dealing with
antisymmetric solutions.

\subsection{Basic solutions: the ZKBP and dipole solutions}
\label{sec:ZKBP_and_dipole}
The lowest-order (i.e., fundamental) solutions to the ambipolar diffusion
equation are particularly simple. The
symmetric solution is
\begin{equation}\label{ZKBP}
\Bz(x,t)= \frac{B_{00}}{s_Z^{}(t)^{1/4}}
\,\left\{1-\left[\frac{x}{L_{00}\, s_Z^{}(t)^{1/4}}\right]^2\right\}^{1/2}, 
\end{equation}
where  $s_z^{}(t)\hskip -3pt = \hskip -3pt1\,+\,4\,t / t_0\,$ and 
\hskip 2pt $t_0 = L_{00}^2 / (\chi_a\,B_{00}^2)$. Here $B_{00}$ and $L_{00}$
represent the field strength at $x=0$ and the spatial domain of the solution,
respectively, both at $t=0$. This is the basic solution for the general
problem of nonlinear diffusion in the particular case that the diffusion
coefficient depends on the square of the variable being diffused; it is
therefore the particular version for ambipolar diffusion of the fundamental
Zeldovich-Kompaneets solution \citep{Zeldovich_Kompaneets_1950,
    Barenblatt1952, Pattle1959}, also known in the mathematical literature as
  the Barenblatt solution. For simplicity, (\ref{ZKBP}) will be called
  the ZKBP solution in the following.  It is plotted, for
$x>0$, as the black curve in Fig.~\ref{fig:the_cartesian_solutions}.

In contrast, the lowest-order antisymmetric solution is
\begin{equation}\label{dipolesol}
\Bz(x,t)=\frac{B_{00}}{s_d(t)^{1/3}}\;
\left\{1-
\left[\frac{x}{L_{00}\,s_d(t)^{1/6}}\right]^{4/3}\right\}^{1/2}
\; \left[\frac{ 6^{3/4}\,x}{L_{00}\,s_d(t)^{1/6}}\right]^{1/3} 
\,,
\end{equation}
where $s_d(t) = 1+4\sqrt{6}\,t/t_0$, while $B_{00}$, $L_{00}$ and $t_0$
have the same meaning as above. 
It is known as the {\it dipole\ solution} and was found by
  \citet{Zeldovich_Barenblatt_1957} (see also \citealt{Zeldovich_book1967});
it is plotted,  for $x>0$, as the blue curve in Fig.~\ref{fig:antisymm}. 
It possesses two lobes (for $x<0$ and $x>0$) that touch at the origin with an $x^{1/3}$ profile. The ambipolar diffusive flux vanishes at the left and right end-points ($x=\pm L_{00}\,(1+4\sqrt{6}\,t/t_0)^{1/6}$), 
but it does not vanish at the origin, so that, mathematically, flux
can diffuse through the origin and cancel with flux from the opposite lobe.  
This is an example of the case discussed in section~\ref{sec:nulls} for pure
ambipolar diffusion causing magnetic flux to cancel, with $\Bz$ vanishing but
having an infinite gradient, although in fact the infinite gradient
  will be resolved by means of an inner Ohmic diffusion layer, as discussed
in that section.

\subsection{The general self-similar formulation} 
\label{sec:self_similar_formulation}
The particular ambipolar solutions (\ref{ZKBP}) and (\ref{dipolesol}) are both of so-called "self-similar" form, namely, the variable $x$ appears only in the combination $x$ divided by a function of $t$, so here we seek more general self-similar ambipolar solutions in that form, namely,
\begin{equation}\label{selfsim}
\Bz(x,t)=\myBzero(t)f(\xi),\ \ \ \ {\rm where} \ \ \ \    \xi(x,t)=\frac{x}{\myL(t)},
\end{equation}
with $f$ a function of the single variable $\xi$, and $\myBzero(t)$ and $\myL(t)$ a time-dependent field strength and a length scale, 
which are all to be determined  as part of the solution. Then the purely
ambipolar diffusion equation reduces to
\begin{equation} \label{eq:after_selfsimilar_2}
0 = - \left(\frac{\myL}{\myBzero}\right)^2 \frac{\dot \myBzero}{\myBzero}\,f
\,+\, \left(\frac{\myL}{\myBzero}\right)^2 \frac{\dot \myL}{\myL}
\,\xi \,f^\prime \,+\, \chi_a\,\left(\strut
f^2\,f^\prime\right)^\prime\;,
\end{equation}
where a dot and a prime
denote ordinary derivatives with respect to $t$ and $\xi$, respectively.

The resulting ambipolar diffusive flux (or rate of transport of magnetic flux) is given by Eq.~(\ref{eq:diff_flux_definition}) with $\eta=0$
and so becomes
\begin{equation} \label{eq:diffusive_flux_separation}
\diffflux = \chi_a\, \frac{\myBzero^3}{\myL}\; \difffluxnodim \;,
\quad\hbox{with}\quad \difffluxnodim\, 
{{\buildrel \rm def \over =}} - f^2 f'.
\end{equation}
We normalize $\myBzero$ and $\myL$ with respect to values $B_{00}$ and $L_{00}$, respectively,  at an  initial time, $t_{00}$, say,   while the unit for time is set as $t_0=L_{00}^2/(\chi_a\,B_{00}^2)$, which is equivalent to putting $\chi_a=1$ in (\ref{eq:after_selfsimilar_2}). 

Then the condition that the coefficients of $f$ and $\xi f^{\prime}$ in Eq.~(\ref{eq:after_selfsimilar_2}) be constants $K$ and $H$, respectively, gives
\begin{equation}\label{eq:K_def}
- \left(\frac{\myL}{\myBzero}\right)^2 \frac{\dot \myBzero}{\myBzero}= -K \;, \qquad 
\;
\left(\frac{\myL}{\myBzero}\right)^2 \frac{\dot \myL}{\myL}=\;H\;.
\end{equation}
The solutions of these equations are

\begin{equation}
\label{eq:B_R_constraint}
\myBzero(t) \; =\; \left[\tauselfsimilarexpansion(t)\right]^{\hskip 2pt \alpha}\;,\quad
\myL(t) \; =\; \left[\tauselfsimilarexpansion(t)\right]^{\hskip 2pt \gamma}\;,
\end{equation}
where
\begin{equation}\label{eq:tau_s}
\tauselfsimilarexpansion(t) =
1 \,+\, 2 \,(H+K)\, (t-t_{00})\;,
\end{equation}
and
\begin{equation}\label{eq:exponents}
\alpha =
\frac{-K}{ 2 K + 2 H} \;, \quad \gamma =
\frac{H}{2 K + 2 H} \;.
\end{equation}

Our self-similar assumption has therefore reduced the partial differential ambipolar diffusion equation to an ordinary differential equation for $f(\xi)$, namely,
\begin{equation}\label{eq:selfsimilar_0}
  0\;=\; K\,f \,+\, H\,\xi\,f^\prime \,+\, \left(
  f^2\,f^\prime\right)^\prime \;;
\end{equation}
\noindent similar equations are found in the applied mathematics
literature \citep{Vazquez_2007, Hulshof_1991}; it is our basic equation and
it needs to be solved with  boundary conditions that differ between the
symmetric and antisymmetric cases.

\subsection{The symmetric case}
\label{sec:symmetric}
\subsubsection{Boundary conditions} 
\label{sec:symmetric_bcs}   

At the origin, we impose
\begin{equation}\label{eq:bcs_0}
\left\{\;
\begin{matrix}
  f &=& 1\\
\noalign{\vspace{2mm}}
  f^\prime &=& 0
  \end{matrix}\;\right\} \quad \hbox{at} \quad \xi=0\;, 
\end{equation}
so, in this case, $\myBzero(t)$ is simply the magnetic field at the origin.
Similarly to what was shown in detail in our previous paper
  (see Fig.~1 in \citealt{Moreno-Insertis_etal_2022} and associated text), a
  continuum of solutions can be found 
  for Eq.~(\ref{eq:selfsimilar_0}) which, in general, extend to
  infinity. Yet, we are interested in solutions with compact support; to
  obtain them, we also impose two conditions at an outer boundary, namely,
\begin{equation} \label{eq:bcs_d}
\left\{\;
\begin{matrix}
  f &=& 0\\
\noalign{\vspace{2mm}}
  (f^3)^\prime &=& 0
  \end{matrix}\;\right\} \quad \hbox{at} \quad \xi=1\;. 
\end{equation}
The first of these is natural, but the second needs explanation. 
If $B(x,t)$ did not change sign in the whole domain, it would be natural to impose constancy of magnetic flux, namely, 
\begin{equation}\label{eq:flux_self}
  \Phi = \int_0^\myL \Bz(x,t)\, dx =\myBzero(t)\,\myL(t) \int_0^1 f(\xi)\,d\xi=\hbox{const}\;,
\end{equation}
which leads to $\myBzero\, \myL = \mathrm{const}$,  $K/H = 1$, and therefore the ZKBP solution. However, we seek instead new self-similar solutions, namely, higher harmonics that possess zeros. In this case we impose instead
\begin{equation}\label{eq:flux_balance}
\int_0^1 f(\xi)\,d\xi=0\;,
\end{equation} 
which leads, after using
Eqs.~(\ref{eq:selfsimilar_0}) and (\ref{eq:bcs_0}),
to
\begin{equation}\label{eq:boundary_condition_edge_1}
(f^3)' \to 0 \quad\hbox{as}\quad \xi \to 1\;.
\end{equation}
Physically, this corresponds to the vanishing of
 the diffusive flux at the outer expanding boundary
so that the magnetic flux is confined in the domain. 

\subsubsection{Solutions}
\label{sec:symmetric_solutions}

In order to find the eigenfunctions and eigenvalues for
Eq.~(\ref{eq:selfsimilar_0}), we have adopted methods and techniques like
those of \cite{Moreno-Insertis_etal_2022}, using, in particular, a stretched
variable $\xitilde$, defined as $\xitilde = H^{1/2} \xi$, and the coefficient
$\Ktilde = K/H$, so that Eq.~(\ref{eq:selfsimilar_0}) reduces to

\begin{equation}\label{eq:selfsimilar_1_cartesian}
  0\;=\; \Ktilde \,f \,+\, \xitilde\,\fracdps{df}{d\xitilde}\,+\,
  \fracdps{d}{d\xitilde}\left(f^2\,\fracdps{df}{d\xitilde} \right) \;.
\end{equation}

\begin{figure}[ht]
\centerline{
\includegraphics[width=\columnwidth]{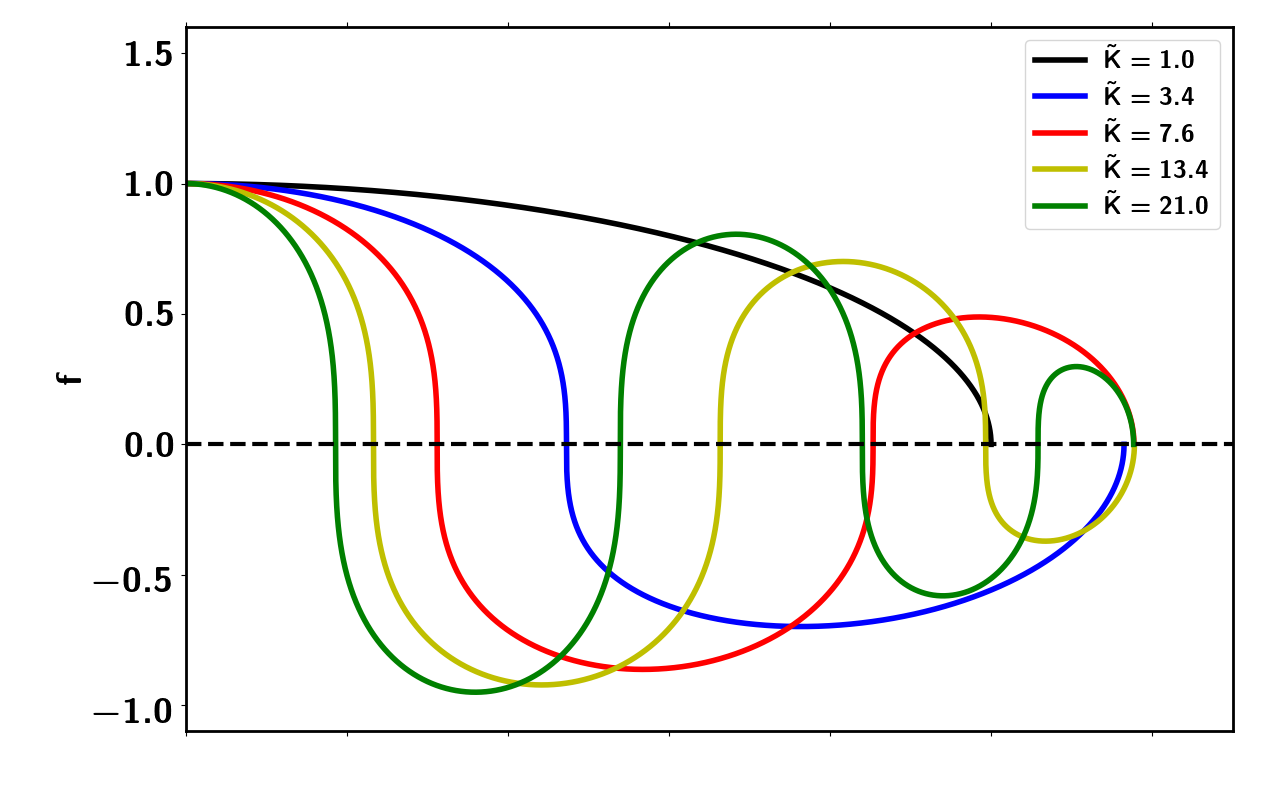}}
\vskip -4mm \hbox to \hsize{\hfill
  \includegraphics[width=0.95\columnwidth]{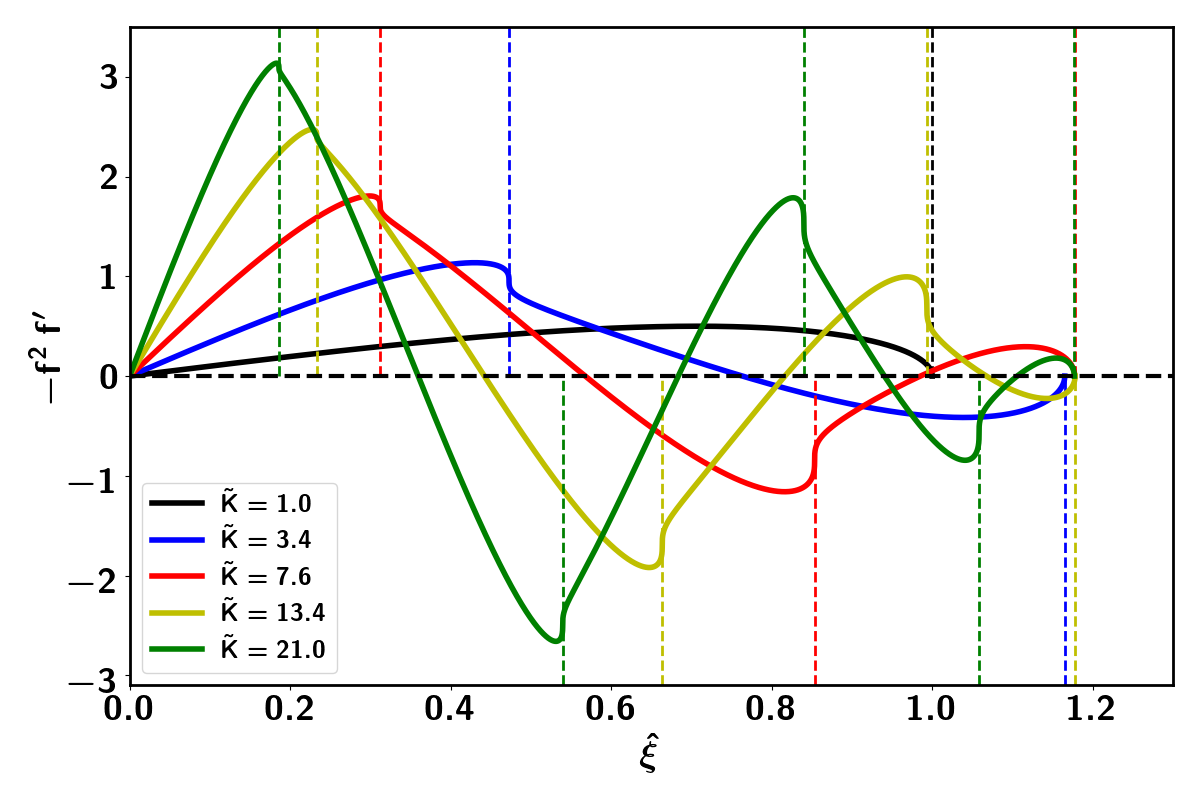}}
\caption{(Top) The fundamental (i.e., ZKBP) and first
  four symmetric harmonics as functions of the stretched
  coordinate $\xitilde = H^{1/2}\,\xi $. The eigenvalues $\Ktilde=K/H$ are
  indicated.
(Bottom) Spatial profile of the dimensionless ambipolar diffusive flux 
  for the eigenfunctions shown in the top panel. The locations of the null
  crossings of the solution are marked with vertical dashed lines.  }
\label{fig:the_cartesian_solutions} 
 \end{figure}

The boundary conditions (\ref{eq:bcs_d})
  serve as quantization rules, so that one finds a countably infinite series
  of solutions for individual values of $\Ktilde$, i.e., 
a fundamental mode and an infinite sequence of harmonics, with each
  of them having one more interior null than the preceding one. We will
  also call these solutions the symmetric {\it eigenmodes} of the
  self-similar problem and the corresponding value of $\Ktilde$ the {\it
    eigenvalue: the term eigenmode and harmonic will be used interchangeably
    in the rest of the manuscript.}  Fig.~\ref{fig:the_cartesian_solutions}
presents the fundamental and first four harmonics (top panel), together with
the corresponding diffusive flux (bottom panel).  The value of $\Ktilde$ for
the harmonics is given in the bottom panel of the figure; 
$\Ktilde$ and further quantities of interest for the fundamental and the
    first four harmonics can be found in Table~\ref{tab:one}.  The fundamental
(or ZKBP) solution is simply $(1-\xitilde^2)^{1/2}$ with $\Ktilde=1$.
The general physical and mathematical behaviour of all solutions is
similar to that of the cylindrical problem presented by \citet{
    Moreno-Insertis_etal_2022}.  Physically, these self-similar solutions describe
pure ambipolar diffusion of a magnetic field possessing a maximum at the
centre ($x=0$) of a Cartesian slab with an initial profile $\Bz(x,0)=f(x)$.
  The effect of ambipolar diffusion is to make the maximum ($\myBzero$)
decrease like a power law in time and the length-scale $\myL$ to increase, with
exponents that depend on $\Ktilde$.
In the process, the outer boundary moves outwards, the integrated
  magnetic flux remains equal to zero while the unsigned flux (i.e., $\int_0^1
  |f(\xi)|\, d\xi$) decreases. The exception to that behaviour occurs for the
  fundamental mode, which has conserved values for both the total and unsigned
  flux. The fundamental mode vanishes only at the outer boundary, while
harmonics of increasing order $n$ vanish at a finite number ($n$) of
intermediate points.  
The passage through the internal nulls is via a singularity that decreases
like $|\hskip 1pt\delta\hskip 1pt |^{1/3}$ with distance, $\delta$, from the
null, whereas, at the outer boundary, the diffusive flux vanishes and the
solution behaves like $|\,2\,H\,\delta\,|^{1/2}$ with distance from the
boundary.  As the magnitude of the field strength decreases in time, magnetic
flux is carried by the ambipolar diffusive flux
(Eq.~\ref{eq:diff_flux_definition}) from the maxima and minima towards the
zeros, where the magnetic flux is annihilated. In practice, the presence of a
small Ohmic diffusivity will resolve the $|\hskip 1pt \delta\hskip
1pt|^{1/3}$ singularities in narrow Ohmic layers, as discussed in
Section~\ref{sec:nulls}.  The same process is present in the antisymmetric
solutions of the next subsection.

\begin{table}[b]
\caption{\label{tab:one} Basic parameters for the first five eigenmodes of
  the symmetric problem.}
\centering
\begin{tabular}{||c||c|c|c|c|c|c||}  
\hline
\rule{0pt}{2.5ex}
\vtop to 3mm{\vfill}& $\Ktilde$  & $\xitildejunction-1 $ & $H -1$ & $K$ & $\alpha$ & $\gamma$ \\
\hline
0& $1.00$ &  $0.000$  &  $0.000$ &   $1.00$  & -0.250 & 0.250 \\ \hline
1& $3.44$ &  $0.165$  &  $0.358$ &   $4.67$  & -0.387 & 0.113 \\ \hline
2& $7.59$ &  $0.177$  &  $0.386$ &   $10.5$  & -0.442 & 0.0582 \\ \hline
3& $13.4$ &  $0.178$  &  $0.388$ &   $18.6$  & -0.465 & 0.0347 \\ \hline
4& $21.0$ &  $0.177$  &  $0.386$ &   $29.1$  & -0.477 & 0.0228 \\ \hline
\end{tabular}
\tablefoot{
The successive columns show the eigenvalues ($\Ktilde = K/H$) and
  the locations of the end points, $\xitildejunction$, of the solutions; 
the corresponding values of $H$ and $K$; and the exponents $\alpha$ and
$\gamma$ of the power laws \hbox{(\ref{eq:B_R_constraint}) --
  (\ref{eq:exponents})}.}  
\end{table}

\subsection{The antisymmetric case}\label{sec:antisymmetric}

\vspace{3mm} In order to find antisymmetric solutions to
Eq.~(\ref{eq:selfsimilar_0}), or, equivalently, after defining again
$\xitilde=H^{1/2} \,\xi$ and $\Ktilde = K/H$, to
Eq.~(\ref{eq:selfsimilar_1_cartesian}), we need to adopt different boundary
conditions and also to reinterpret what we mean by the function
$\myBzero(t)$.

\subsubsection{Equations and boundary conditions}\label{sec:antisymmetric_eqs}

A set of boundary conditions at the origin $\xitilde=0$ appropriate for 
  antisymmetric solutions is: 
\begin{equation}\label{eq:bcs_antisym}
\left\{\;
\begin{matrix}
  f &=& 0\\
\noalign{\vspace{2mm}}
   -f^2\,\fracdps{df}{d\xitilde} &=& -1
  \end{matrix}\;\right\} \quad \hbox{for} \quad \xitilde \to 0\;. 
\end{equation}

For the second of these, the value $-1$ is chosen for simplicity, but any
other non-zero (finite) value could be imposed instead. Irrespective of the
actual value chosen, this condition implies that $f(\xitilde) \propto
\xitilde^{1/3}$ as $\xitilde \to 0$. Hence, $f$ is infinitely steep at the
origin, and this family of solutions possesses a singularity there. 
As for the symmetric case, imposing only the boundary condition at
  the origin one can find a continuum of solutions which, in most cases,
  extend to infinity. To extract a set of eigenmodes, i.e., solutions with
  compact support, an outer boundary condition at a finite location must be
  imposed; yet, the flux condition and the boundary condition must now be
  reconsidered for the antisymmetric case.
The integration of an antisymmetric solution along the whole
real axis naturally yields zero, so that the net magnetic flux automatically
vanishes, and the constraint of flux conservation does not lead in this case
to a condition for the flux integral on each semi-axis ($\xitilde>0$ or
$\xitilde<0$). In contrast to the previous cases, $\int_0^1 f(\xitilde)
\,d\xitilde$ can now be different from zero for any of the harmonics without
violating the condition of flux conservation.  We realise, however, that the
original boundary conditions (Eq.~\ref{eq:bcs_d}) at the outer edge must be
retained here as well: (a) the search for solutions with compact support
leads to the first of those conditions; (b) also, imposing the second
of Eq.~(\ref{eq:bcs_d}) is just requesting that there is no flux loss nor
gain through the outer boundary. This is a reasonable condition to apply not
just on mathematical grounds but also through the simple physical
consideration that no magnetic flux can be gained from (or lost to) an
external unmagnetized domain. So, Eq.~(\ref{eq:bcs_d}) will be used here as
outer boundary conditions to define the eigenvalue problem for the
antisymmetric case as well. To complete the description, one needs to revisit
the normalization conditions and the evolutionary power laws used for the
symmetric case, as follows.

So far, we have just obtained solutions for the eigenvalue problem
  Eq.~(\ref{eq:selfsimilar_1_cartesian}) subject to the boundary conditions
  appropriate to the antisymmetric case, namely Eqs.~(\ref{eq:bcs_d}) and
  (\ref{eq:bcs_antisym}). However,   $\myBzero(t)$
  can no longer be determined from the value of the 
  magnetic field at $\xitilde=0$ (which was our choice for the symmetric
  case), but  is here instead dictated by the boundary condition $f^2 df/d\xitilde \to 1$ 
 as $\xitilde\to 0$. Using the definition of diffusive flux,
 Eq.~(\ref{eq:diff_flux_definition}) with $\eta=0$ and 
Eq.~(\ref{eq:diffusive_flux_separation}), and calling $(\diffflux)_0^{}(t) =
\diffflux(x\to 0, t)$, we conclude that $\myBzero(t) = [-(\diffflux)_0^{}
  (t)\,\chi_a\,\myL(t)/H^{1/2}]^{1/3},$ with $B_{00}$ still given by
$\myBzero(t_{00})$.
The corresponding choice of $L_{00}^2/(\chi_a\,B_{00}^2)$ as a 
  time unit is equivalent, as in Section~\ref{sec:self_similar_formulation}, to choosing $\chi_a=1$ in Eq.~(\ref{eq:after_selfsimilar_2})
  and leads to the same equations (\ref{eq:K_def}) -- (\ref{eq:tau_s}) for
  the time-dependence, which contain the essential power laws of the problem.   

\subsubsection{Solutions}\label{sec:antisymmetric_solutions}

\begin{figure}[ht]
\centerline{\includegraphics[width=\columnwidth]{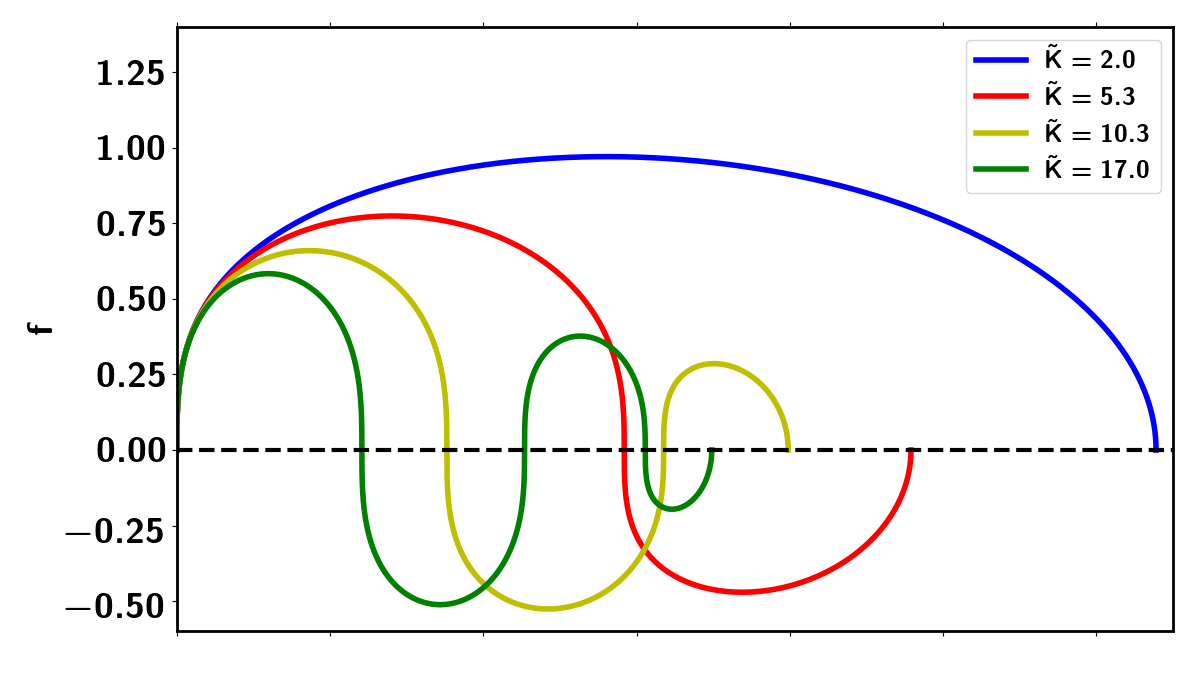}}
\vskip -4mm \hbox to \hsize{\hfill
  \includegraphics[width=0.985\columnwidth]{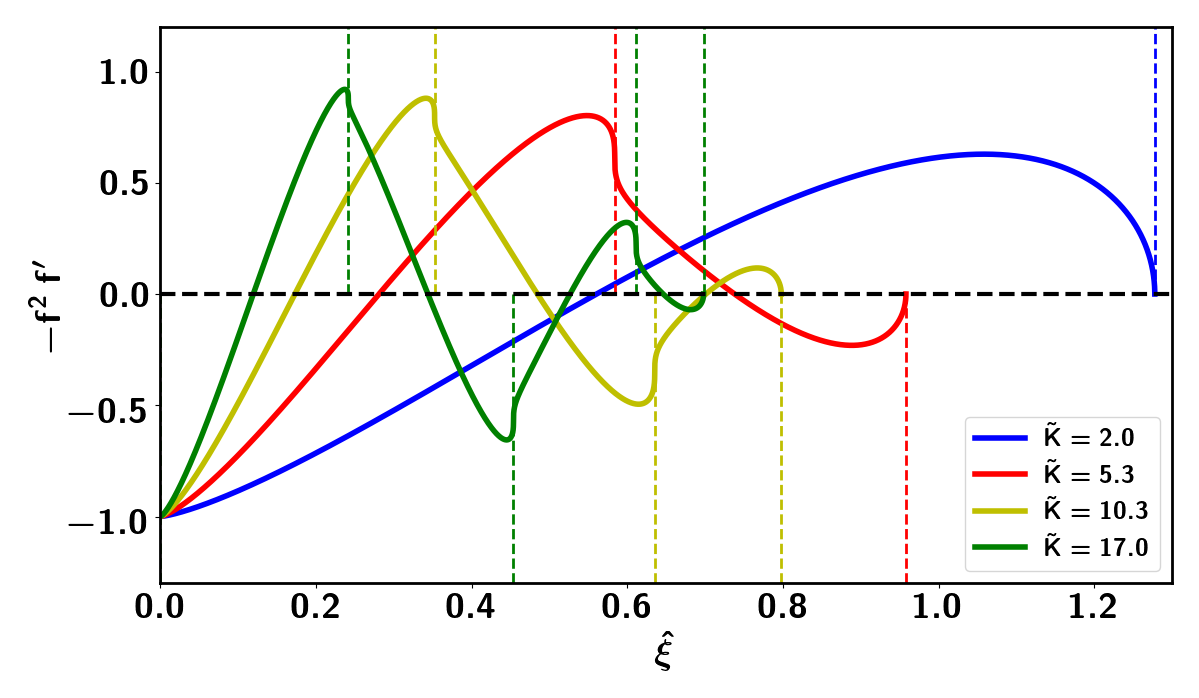}}
\caption{(Top) The fundamental (i.e., dipole) and first
  three antisymmetric harmonics as functions
  of the stretched   coordinate $\xitilde = H^{1/2}\,\xi $. The eigenvalues
  $\Ktilde=K/H$ are   indicated.
(Bottom) Spatial profile of the dimensionless diffusive flux $\difffluxnodim$
  for the eigenfunctions shown in the top panel. The locations of the null\href{}{}
  crossings of the solution are marked with vertical dashed lines.  }
\label{fig:antisymm} 
 \end{figure}

Fig.~\ref{fig:antisymm} shows the solutions for the first four antisymmetric
harmonics. Here, as  previously, the term \textit{n-harmonic} refers to a 
solution with $n$ zeros in the $\xitilde \ge 0$ domain, excluding the outer
front.   
The first harmonic has an exact analytical expression 
\citep[see][]{Zeldovich_Barenblatt_1957, Bernis_etal_1993}, namely:
\begin{equation}\label{eq:antisymmetric_analytical}
f(\xitilde) = \xitilde^{1/3}\;\left(C - \frac{3}{2}\,\xitilde^{4/3}\right)^{1/2}\;.
\end{equation}
To match the boundary condition for the derivative at the centre
(Eq.~\ref{eq:bcs_antisym}, second line), one has to choose $C=3^{2/3}$, 
so that  
 \begin{equation} \label{eq:1stharmonic}
f(\xitilde) = (3\,\xitilde\hskip 1pt)^{1/3}\;\left(1 -
\frac{3^{1/3}}{2}\,\xitilde^{4/3}\right)^{1/2}\;.
\end{equation} 
We have already come across this solution (Eq.~\ref{dipolesol}),
  but now we have expressed it in terms of the self-similar variables. 
As expected, all solutions have infinite slope at the origin, so their numerical integration has to be started using a phase-plane
analysis \citep{Moreno-Insertis_etal_2022}. The eigenvalue 
$\Ktilde$ and other parameters are given in Table~\ref{tab:two}. Among other things we see that the values for $\Ktilde$
for the nth-harmonic here are intermediate between those for the $n$ and
$n-1$ harmonics for the symmetric case (see Table~\ref{tab:one}), in
agreement with the results of \citet{Hulshof_1991}. The first harmonic 
constitutes a particularly good test for MHD codes, since it has an analytical
solution and contains a singularity at the origin of type $f
\propto \xitilde^{1/3}$, with non-zero diffusive flux, $\difffluxnodim \ne  0$ (Eq.~\ref{eq:diffusive_flux_separation}). 

\begin{table}[b]
\caption{\label{tab:two} Basic parameters for the first four eigenmodes of
  the antisymmetric problem.}
\centering
\begin{tabular}{||c||c|c|c|c|c|c||}  
\hline
\rule{0pt}{2.5ex}
\vtop to 3mm{\vfill}& $\Ktilde$  & $\xitildejunction$ & $H$ & $K$ & $\alpha$ & $\gamma$ \\
\hline
1& $2.00$ &  $1.278$  &  $1.633$ &  $3.27$  & -0.333 & 0.167 \\ \hline
2& $5.30$ &  $0.958$  &  $0.918$ &  $4.87$  & -0.421 & 0.0793 \\ \hline
3& $10.3$ &  $0.798$  &  $0.637$ &  $6.56$  & -0.456 & 0.0443 \\ \hline
4& $17.0$ &  $0.698$  &  $0.487$ &  $8.28$  & -0.472 & 0.0278 \\ \hline
\end{tabular}
\tablefoot{
The successive columns show the eigenvalues ($\Ktilde = K/H$) and
  the locations of the end points, $\xitildejunction$, of the solutions; 
the corresponding values of $H$ and $K$; and the exponents $\alpha$ and
$\gamma$ of the power laws \hbox{(\ref{eq:B_R_constraint}) --
  (\ref{eq:exponents})}.}  
\end{table}

\section{Mode stability: a few time-dependent calculations}
\label{sec:solutions_overview}

In this section, we briefly present a suite of time-dependent numerical
solutions of the ambipolar diffusion equation,
Eqs.~(\ref{eq:ad_diffusion_equation})-(\ref{eq:diff_flux_definition}) with
$\eta=0$, with various initial conditions, to obtain empirical hints on
whether the eigenmodes are stable, in the sense that the numerically
calculated time-evolution converges to that eigenmode or not. To that end we
use a simple integrator based on the method of lines with second-order
spatial derivatives and first-order update in time. This integrator keeps the
net flux constant during the integration to machine precision, which turns
out to be of interest for the conclusions reached in the following. As seen
below, this integrator accurately keeps the shape and time-dependence of the
fundamental mode and of the first antisymmetric harmonic (which are stable,
as proved in the mathematical literature and explained below) and of the
first symmetric harmonic (which is likely to be stable), so it is a good tool
for the study undertaken in this section.

\subsection{Solutions with a symmetry condition about $x=0$}
\label{sec:solutions_symmetric_bc}
Here the 1D Cartesian case is considered with boundary conditions at the
origin that keep the solution symmetric about $x=0$.  Since all the solutions
have compact support, we set minimally invasive boundary conditions
at the 
right-hand boundary, namely, $f= df/d\xitilde =0$, 
which guarantee that the solutions vanish there until the expanding solutions reach the boundary.

\subsubsection{Universality of the ZKBP solution for initial conditions that are not in flux balance.} \label{sec:ZKBP_universality}

 Whenever the initial conditions are not in flux balance, so that the net
 flux does not vanish, we find that the solution tends asymptotically in time
 to the ZKBP solution for that particular flux. This is expected, since it is
 the relevant stable solution for the general case of flux cancellation with
 unequal amounts of flux in the two polarities
 \citep[see][]{Kamin_Vazquez_1991, Vazquez_2007}. This result applies as
 well when, in the numerical solution, the nonzero net flux is
 caused by rounding errors in the numerical discretization: even if one
 starts with initial conditions in flux balance ($\Phi_{mag}=0$),
discretization of the initial condition leads to a  small nonzero net flux,
typically $3$, $4$, or $5$ orders of magnitude smaller than the initial
unsigned flux. The integrator for these calculations keeps the net flux
constant during the integration; so, the tiny initial net flux is maintained,
and it plays a 
negligible role for as long as the unsigned flux of the physical solution is
orders of magnitude larger. But the unsigned flux of the eigenmodes
continuously decreases in time. Hence, by waiting a sufficiently long time,
the initially tiny net flux will start to play a role, and the final outcome
will be a ZKBP solution.  Yet, this is generally of little physical
interest if the time needed
to reach such a state is very long, much longer than the other
  relevant physical timescales.  In the following, we therefore disregard the tiny net initial flux, at least for as long as the unsigned flux of the solution is well above it.

\subsubsection{Stability of the first symmetric eigenmode.} \label{sec:stability_1st_eigenmode}

\begin{figure}
\centerline{
\hskip -2mm \includegraphics[width=0.51\textwidth]{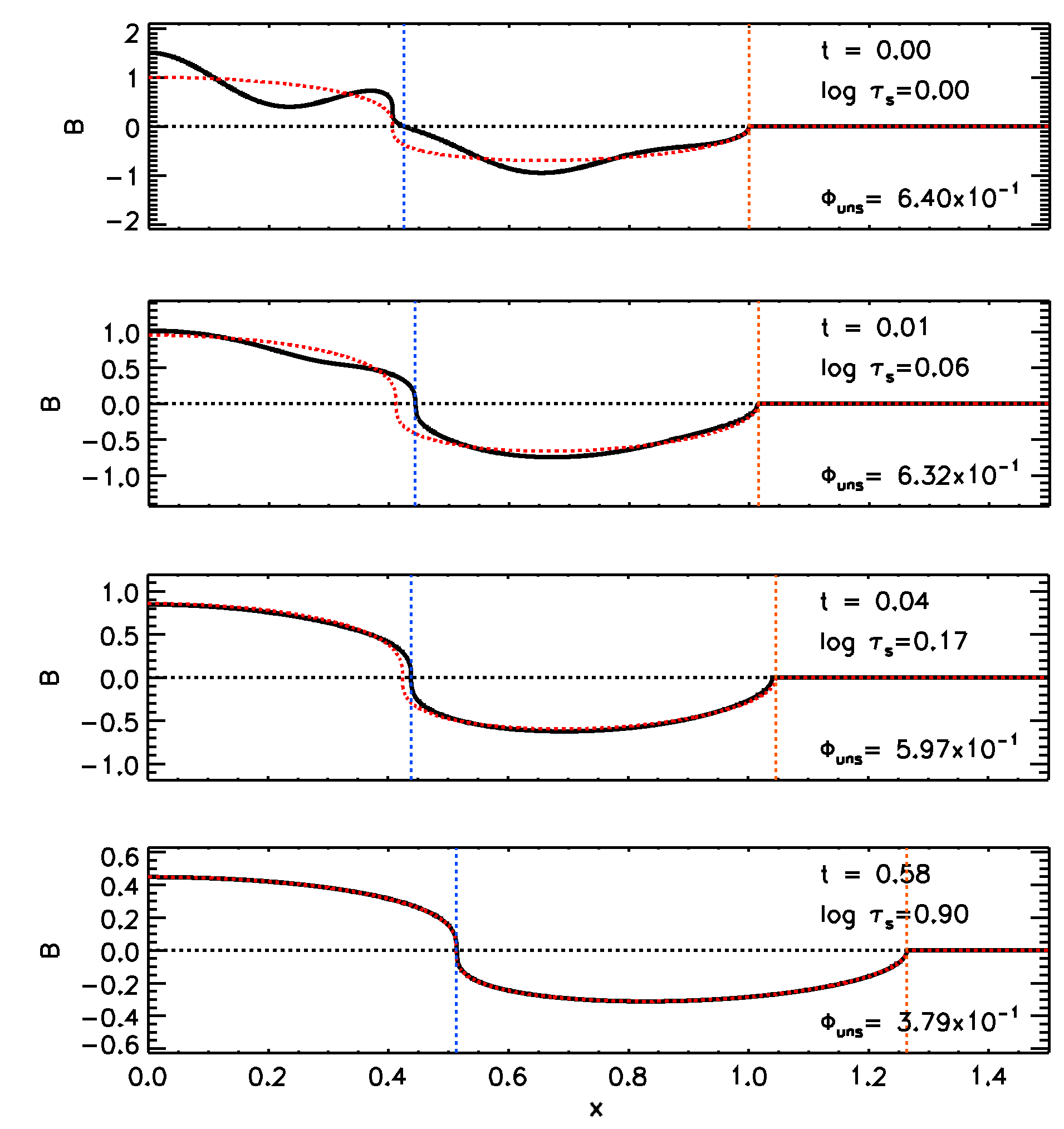}
}
\caption{Magnetic field evolution starting from a superposition of the
  symmetric first harmonic and a zero net-flux finite perturbation
  which does not add any extra internal null. The panels show four
  selected instants (top: initial state; bottom: asymptotic
  evolution). Note the different ranges of the ordinate axes
  in the different panels.  The
  evolving self-similar first symmetric harmonic is shown as a dashed line in
  red. The net magnetic flux is $\Phi_{tot} = \pot{7.7}{-6}$; it does not
  change during the time evolution.  The base-$10$ logarithm of the diffusive
  time is indicated.  Vertical dashed lines: locations of the
    internal zero (blue) and outer front (red) of the solution.  An
  associated animation shows intermediate stages including the
    solution (upper panel) and the corresponding diffusive flux (lower
    panel); the associated movie is available online}
\label{fig:figV4}
\end{figure}

The first symmetric eigenmode is likely to be an attractor for all symmetric
solutions with zero net flux and a single internal null point (for a
  mathematical discussion of various properties of this eigenmode see
  \citealt{Bernis_etal_1993} and \citealt{Hulshof_1991}).  
The experience gained from many numerical solutions carried out in the
context of this paper, although insufficient to prove any general
mathematical properties, is fully compatible with that hypothesis. For
instance, we have tested cases in which, as initial condition at $t=0$, the
first symmetric eigenmode was used with the addition of a zero-flux finite
deviation from it, namely, an ad-hoc deviation that, importantly, does not
add zeros to the initial function. As apparent in Fig.~\ref{fig:figV4} and the
associated animation, the solution quickly converges to the shape of the
first symmetric harmonic and remains there as its asymptotic state.  For the
size and shape of the perturbation used here, by $t=0.21$ ($\log
\tauselfsimilarexpansion = 0.6$), the numerical solution and the exact 
self-similar solution for the first harmonic are virtually indistinguishable
in the diagrams, with the height of both diminishing in time, and their
spatial extent increasing, according to the self-similar power laws
(\ref{eq:B_R_constraint}) -- (\ref{eq:tau_s}).

\subsubsection{Stability of the higher symmetric harmonics.} \label{sec:stability_higher_harmonics}

\begin{figure}
\centerline{\hskip -3mm
\includegraphics[width=1.0\columnwidth]{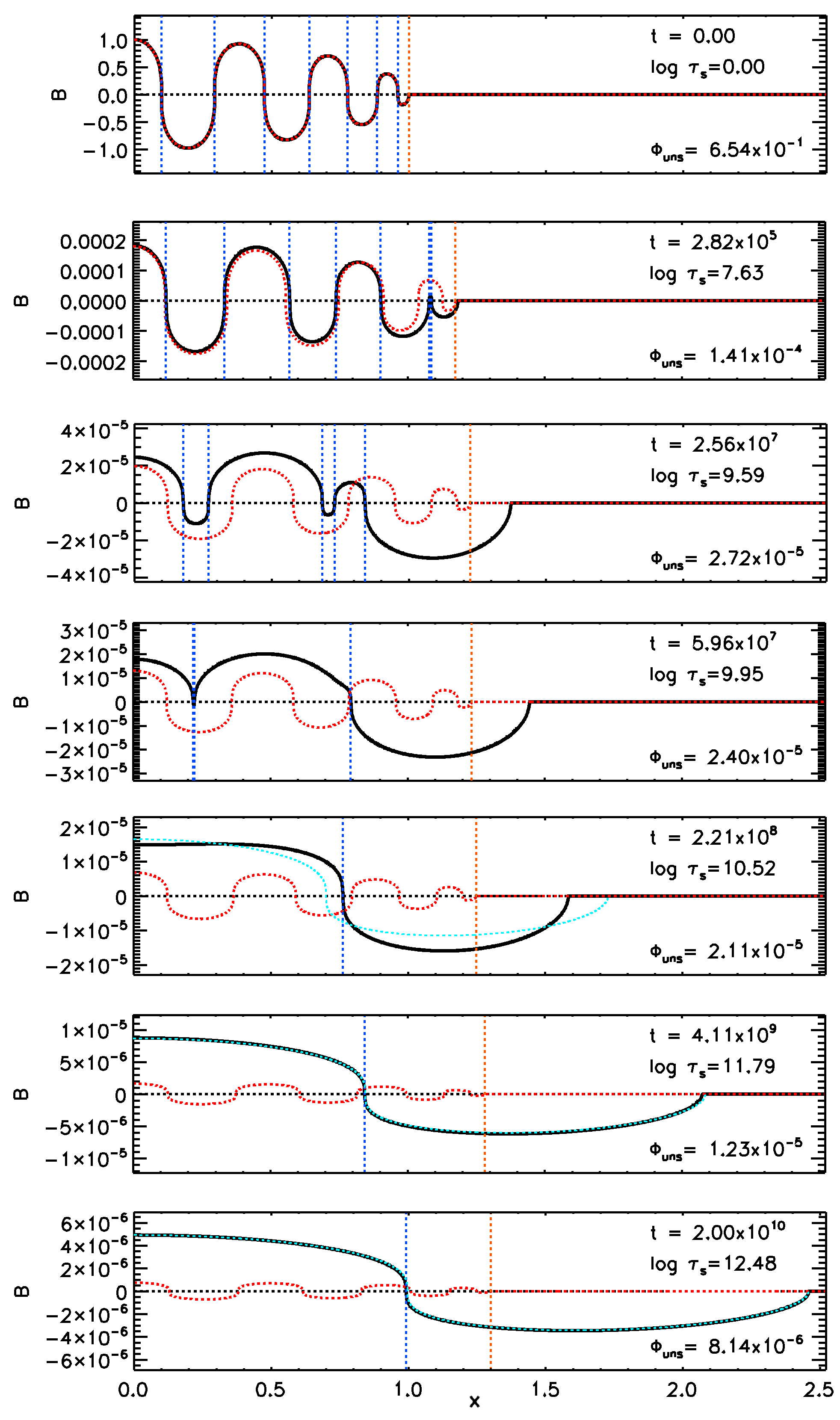}}
\caption{Selected stages in the evolution of the numerical solution of the
  ambipolar diffusion equation with initial condition having the shape of the
  7th symmetric self-similar harmonic and with the numerical mesh containing
  2048 grid points. Red: the exact solution for the 7th self-similar
  harmonic. Black: actual numerical solution. Cyan (only in the
    three lowermost panels): exact solution for the 1st self-similar
  harmonic that coincides with the numerical solution at the final time
  calculated in the figure. Note the very different ranges of the ordinate axes
  in the different panels.
Vertical dashed lines: locations of the internal
  zeros of the numerical solution (blue); outer front of the exact 7th
  symmetric self-similar harmonic (red). The total (or net) magnetic flux is
  $\pot{-7.4}{-8}$.  An associated animation is provided that shows
  intermediate stages including the solution (upper panel) and the
    corresponding diffusive flux (lower panel); the associated movie is
    available online}
\label{fig:figV5}
  \end{figure}

The stability of the higher harmonics does not appear to have been solidly
established in the mathematical literature, but to construct such
mathematical proofs is outside the scope of the present paper. From our
numerical experiments starting with the exact shape of one of the higher
harmonics (say, the 6th or the 7th), and so with zero initial flux apart from
a tiny rounding error, we observe how, when the spatial resolution is high,
they begin by following the exact evolutionary pattern of the harmonic (i.e.,
maintaining the self-similar shape and following the power laws in time and
space corresponding to that harmonic) to high accuracy; however, eventually
those solutions deviate from the exact self-similar shape and evolve toward a
profile with a single internal null, close to the first symmetric harmonic,
with the other internal nulls cancelling in pairs (or being expelled
individually through the outer boundary). An example with intermediate
resolution is shown in the animation accompanying Fig.~\ref{fig:figV5}. The net
magnetic flux (which is caused by the numerical discretization) is
$\pot{-7.4}{-8}$, i.e., tiny in comparison with the initial unsigned flux
$\pot{6.5}{-1}$; the deviation from the original self-similar shape caused by
the lack of stability of the higher harmonic starts to be apparent at roughly
$t\approx \pot{1.5}{3}$, when the unsigned flux has reduced to $\approx
\pot{1.7}{-3}$, still four orders of magnitude larger than the net flux. The
process of transformation into the first symmetric harmonic takes place
starting at that time; to observe that process, we have run the calculation
until an advanced time, namely $t=\pot{2}{10}$, when the transformation was
clearly complete, and at that point fit the shape of a first symmetric
harmonic, for which we know the exact evolution prior and after that time
according to the self-similar power laws of
Section~\ref{sec:self_similar_formulation}. Starting at time
$t=\pot{2.2}{8}$, in the animation and in Fig.~\ref{fig:figV5} we have
superimposed the shape of that exact self-similar first symmetric harmonic
  as a cyan dashed curve.  We can see how the original solution evolves
  toward that harmonic, as if the latter were acting as an
  attractor; in fact, from, roughly, $t=\pot{7}{9}$ onward the two solutions
  are indistinguishable in the figure and animation, keeping an excellent
  fit for the rest of the calculation.

The foregoing can only be taken as a hint of an instability of the higher
harmonics and of the suggestion that the first symmetric harmonic acts as an
attractor for them. A detailed study, with very high spatial resolution and
possibly comparing different numerical methods must be left for future work;
exact mathematical proofs, in turn, would need a sophisticated applied
mathematics treatment.

\subsection{Time evolution with no boundary condition imposed at $x=0$ and
  transforming into the dipole solution}\label{sec:solutions_no_symmetry_bc}

To test empirically for the stability of the first symmetric harmonic subjected to more general perturbations, we extend the initial condition to the negative $x$-axis but maintain the eigenmode shape with even symmetry for the initial condition.  To that shape, perturbations are added of either symmetric or antisymmetric shape to see if the first harmonic is indeed stable.
We first report on simple cases, which need not be graphically illustrated:
(1) If we just let it go with a very small net flux (arising from numerical rounding errors) but otherwise equal to the first symmetric harmonic, it maintains that shape and obeys the corresponding power laws.  (2) If one adds a not necessarily small perturbation with no net flux and which is symmetric about $x=0$, the solution quickly converges to the first symmetric harmonic, following the corresponding power laws.

\begin{figure}
\centerline{
\includegraphics[width=0.5\textwidth]{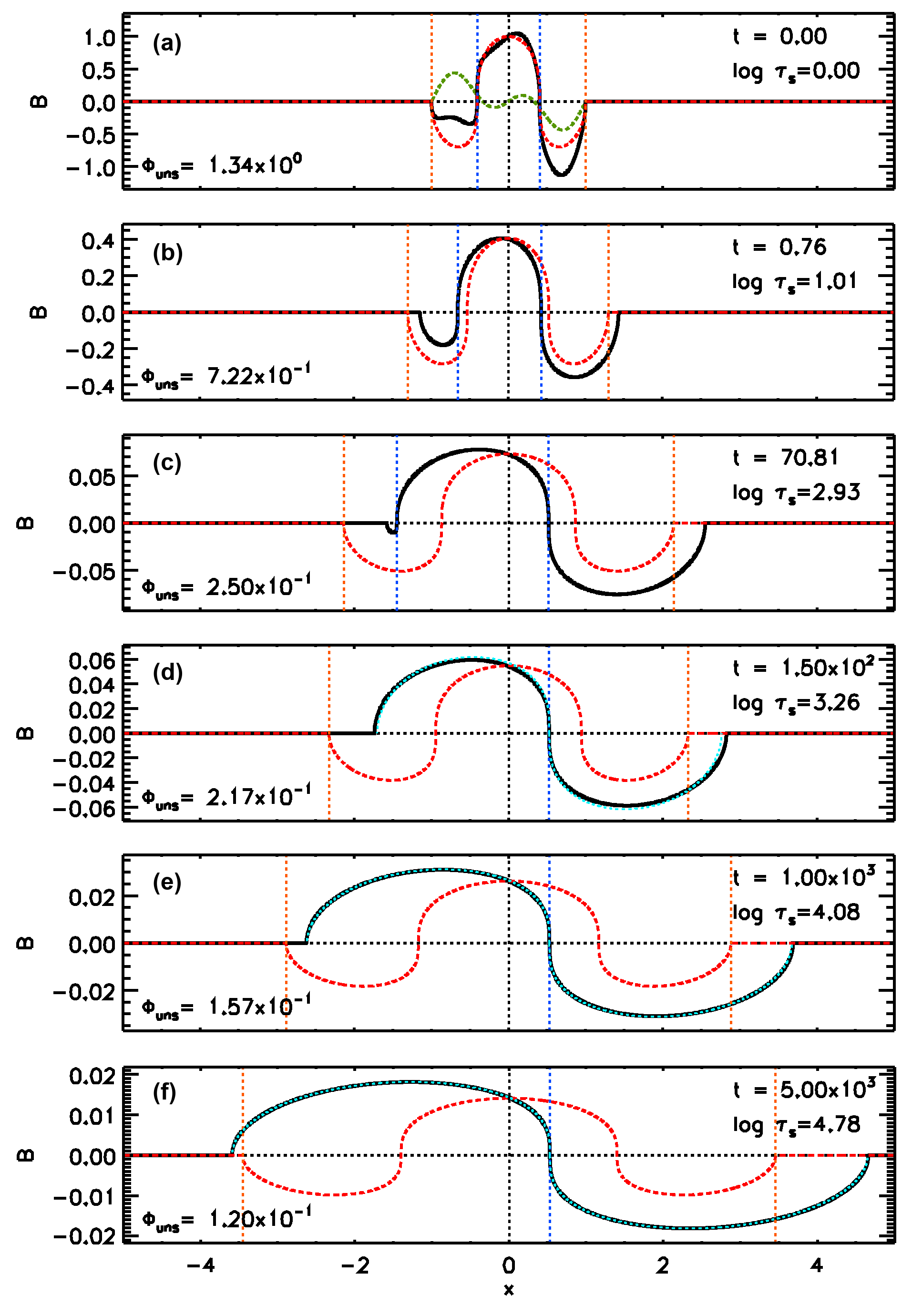}
}
\caption{Selected stages in the evolution of the numerical solution described
  in Sect.~\ref{sec:solutions_no_symmetry_bc}. Panel (a): initial condition
  (black solid curve), resulting from the addition of the exact first
  symmetric self-similar harmonic (red dashed) and a finite antisymmetric
  perturbation of zero net flux (green dashed). In panels (d) - (f), the
  exact self-similar dipole harmonic evolving towards the final shape of the
  numerical solution is superimposed in cyan.  Note the
      different ranges of the ordinate axes in the different panels.  An
  associated animation is provided that shows intermediate stages including
  the solution (upper panel) and the corresponding diffusive flux (lower
  panel); the associated movie is available online.}
\label{fig:sym_to_dipole}
\end{figure}

The next case, (3), is perhaps less obvious and we illustrate it here with a
figure (Fig.~\ref{fig:sym_to_dipole}) and an associated animation. As initial
condition (top panel), we take the same first harmonic symmetric about $x=0$
as in the previous two cases (red dashed curve), but now add a finite
perturbation with an antisymmetric shape (green dashed curve). The solution
in time for the superposition is shown as a solid black curve in all panels
whereas the corresponding solution of the symmetric self-similar
first-harmonic used to build the initial condition is shown as a red dashed
line. It is remarkable to see that the solution quickly loses the internal
null point on the left as it is ejected through the outer front (panels b and
c), and it adopts an antisymmetric shape which eventually behaves in a
self-similar fashion, namely, like the fundamental dipole solution (panels d,
e and f). In the latter three panels, which are separated by a long time
interval (from $\tau_s = \pot{1.5}{2}$ to $\tau_s = \pot{5}{3}$), the
time-dependent evolution of the dipole mode that best fits the final value of
the solution in that time series is added using a cyan dashed line: as can be
seen, at least in the time range $\tau_s = \pot{6}{2}$ to $\tau_s =
\pot{5}{3}$, the actual solution and the dipole one are very close to each
other; this is true also for the flux, as can be checked in the associated
animation.

\section{Tests of the Bifrost code and its ambipolar diffusion solver}
\label{sec:bifrost}

We have conducted numerical experiments in double precision with the 3D
radiation-MHD Bifrost code \citep{Gudiksen_etal_2011} using the ambipolar
diffusion solver developed by
\cite{Nobrega-Siverio_etal_2020a}; the experiments follow a similar procedure
to those in \cite{Moreno-Insertis_etal_2022}.  
In doing so, we shall  refer here to the solutions derived in the earlier sections of this paper as {\it theoretical solutions}, whereas those that are calculated with the Bifrost code will be called {\it Bifrost solutions}.  For brevity, we focus on the symmetric solutions.

\begin{figure*}[ht]
\centerline{
\includegraphics[width=0.37\textwidth]{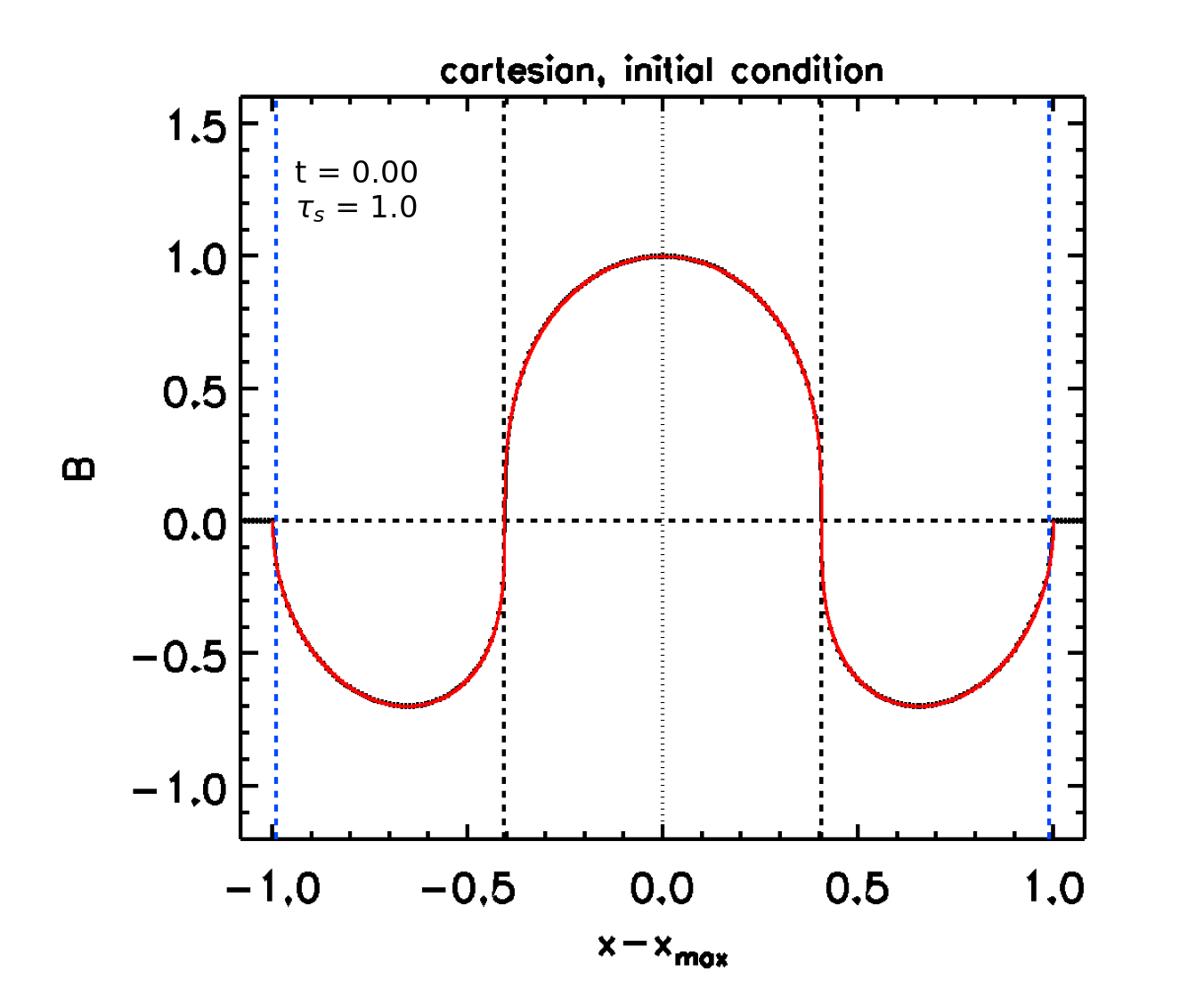}\hskip -4mm
\includegraphics[width=0.37\textwidth]{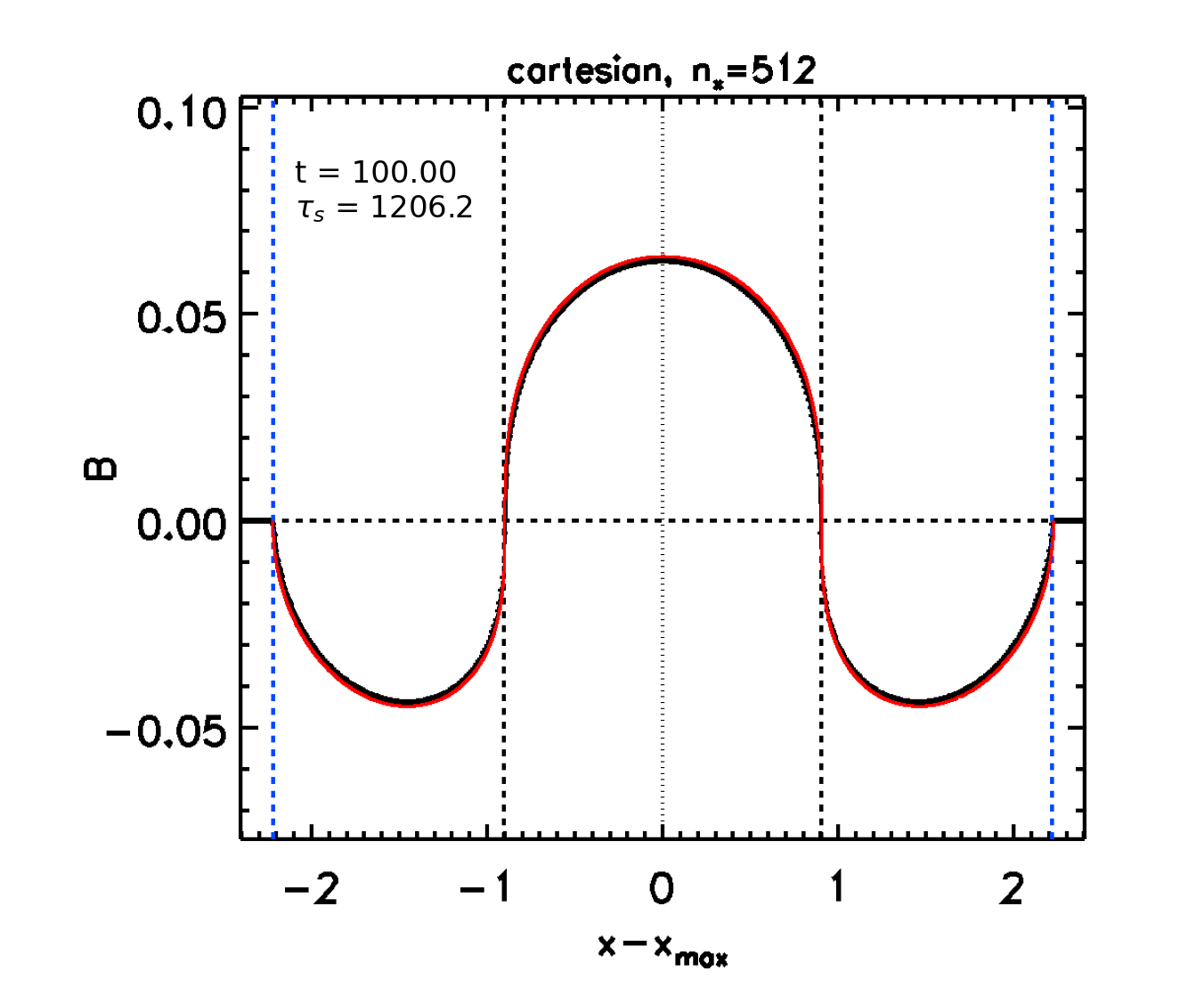}\hskip -4mm
\includegraphics[width=0.36\textwidth]{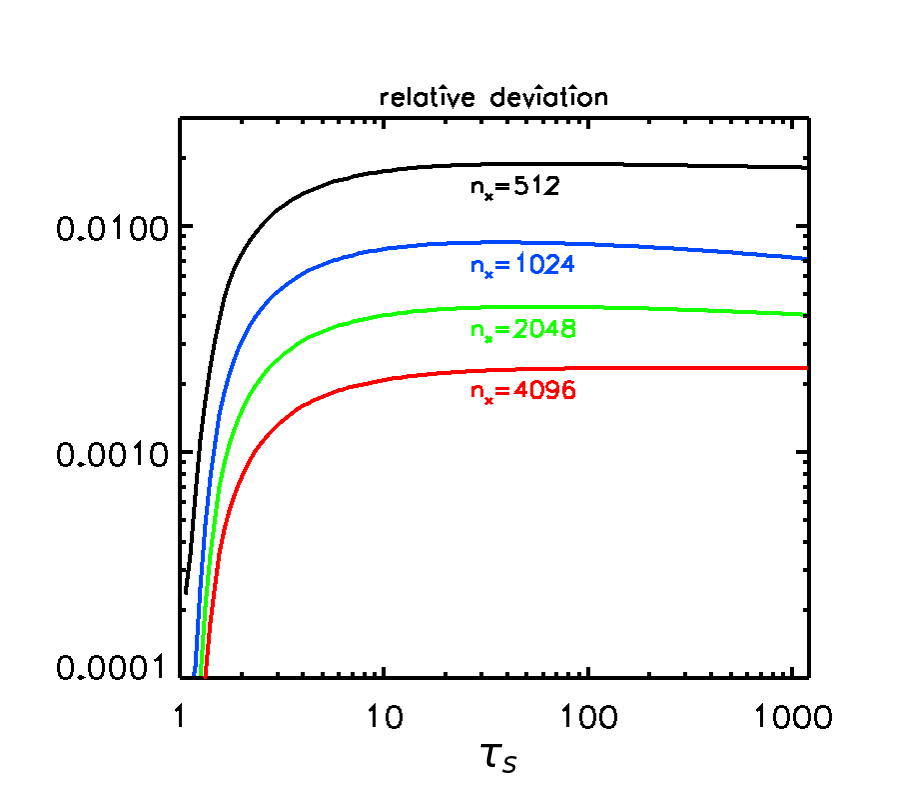}}
\caption{A comparison of the Bifrost runs (black thick curve) with the theoretical solutions (red curves), showing 
  the initial condition (left panel) and the final configuration
  for $n_x=512$ (central panel).  
  The relative mutual deviation of the maxima between the theoretical and Bifrost solutions for four resolution levels is given in the right panel.}
\label{fig:bifrost_cartesian}
\end{figure*}

\begin{figure*}[h]
\centerline{
  \includegraphics[width=1.02\textwidth]{figure_10.png}}
\caption{Time-dependence of the maximum of the magnetic field (left panel), the
  location of the singular points [middle panel; upper curve for
    $\myL(t)$, lower curve for $\myL_{cs}(t)$] and the value of the integral
  $\int B(x,t)\, dx$ in the central lobe of the solution (right panel) for the
  solution of the first harmonic,  illustrated in Fig.~\ref{fig:the_cartesian_solutions}. The Bifrost solution is shown as a black solid line and the theoretical solutions as red dashed lines.}
\label{fig:power_laws_cartesian}
\end{figure*}

As a first step one would have to consider the fundamental ZKBP solution
(Section~\ref{sec:ZKBP_and_dipole}) shown as the black curve of
Fig.~\ref{fig:the_cartesian_solutions}. This is the most stable of the
eigenmodes, and the asymptotic solution for all initial conditions with
non-zero flux; also, importantly for the test, it does not have any internal
singularity. So, we expect the Bifrost calculation to yield satisfactory
results for this solution. Indeed, a first Bifrost test for this eigenmode
was already presented in the paper by 
  \citet{Nobrega-Siverio_etal_2020a} with very good results. A far more
  stringent test for numerical codes is provided by the harmonics to the
  eigenvalue equation (\ref{eq:selfsimilar_0}) (or, equivalently,
  Eq.~\ref{eq:selfsimilar_1_cartesian}) found in Sec.~\ref{sec:self-similar}.  The reason for this is the presence of internal current sheets with a finite diffusive flux across them. Since the magnetic field possesses an infinite gradient there, calculating the
diffusive flux at such singularities is highly challenging for standard
codes, which cannot integrate by going to the phase plane. Here we first show the tests carried out for the first symmetric harmonic
(Section~\ref{sec:bifrost.1}) and then those for one of the higher harmonics (Section~\ref{sec:bifrost.2}).

\subsection{First symmetric harmonic}\label{sec:bifrost.1}

The first symmetric harmonic (Section~\ref{sec:ZKBP_and_dipole}) is
  shown as the blue curve in Fig.~\ref{fig:the_cartesian_solutions}. As
  initial condition for the 
  Bifrost calculation we use that theoretical solution for the right half of
  the domain centred on an arbitrary coordinate $x=(x_{max})_0^{}$, together with its mirror-symmetric counterpart for the left half.
At time $t=0$, we
arbitrarily centre the function at $x=(x_{max})_0^{}=2$ and locate its outer
front at $x=3$
on the right and $x=1$ on the left, in keeping with the normalization
$\myL(0) = 1$ introduced in Section~\ref{sec:self_similar_formulation}; the
function vanishes outside the interval $(1,3)$ in~$x$. The solution is
expected to expand in space according to the rules explained in
Section~\ref{sec:self_similar_formulation}. To cope with that expansion, the integration
domain for the numerical calculation is the interval $(x_L, x_R) = (-0.46,
4.66)$; note that we have set the centre of the function, $(x_{max})_0^{}$, on
purpose at a location not coinciding with the centre of the numerical grid.
At each time-step we calculate the position of the maximum of the numerical
solution, $x_{max} (t) $, and define ${\hat x} = x-x_{max}(t)$. The numerical solution is then compared with the theoretical one for the first harmonic of
Section~\ref{sec:symmetric} using the time-dependence
(\ref{eq:B_R_constraint}) -- (\ref{eq:tau_s}), $\xitilde = H^{1/2}\,\xi$, and
$\xi = {\hat x} / \myL(t)$.

\begin{figure*}
\centerline{\includegraphics[width=0.38\textwidth]{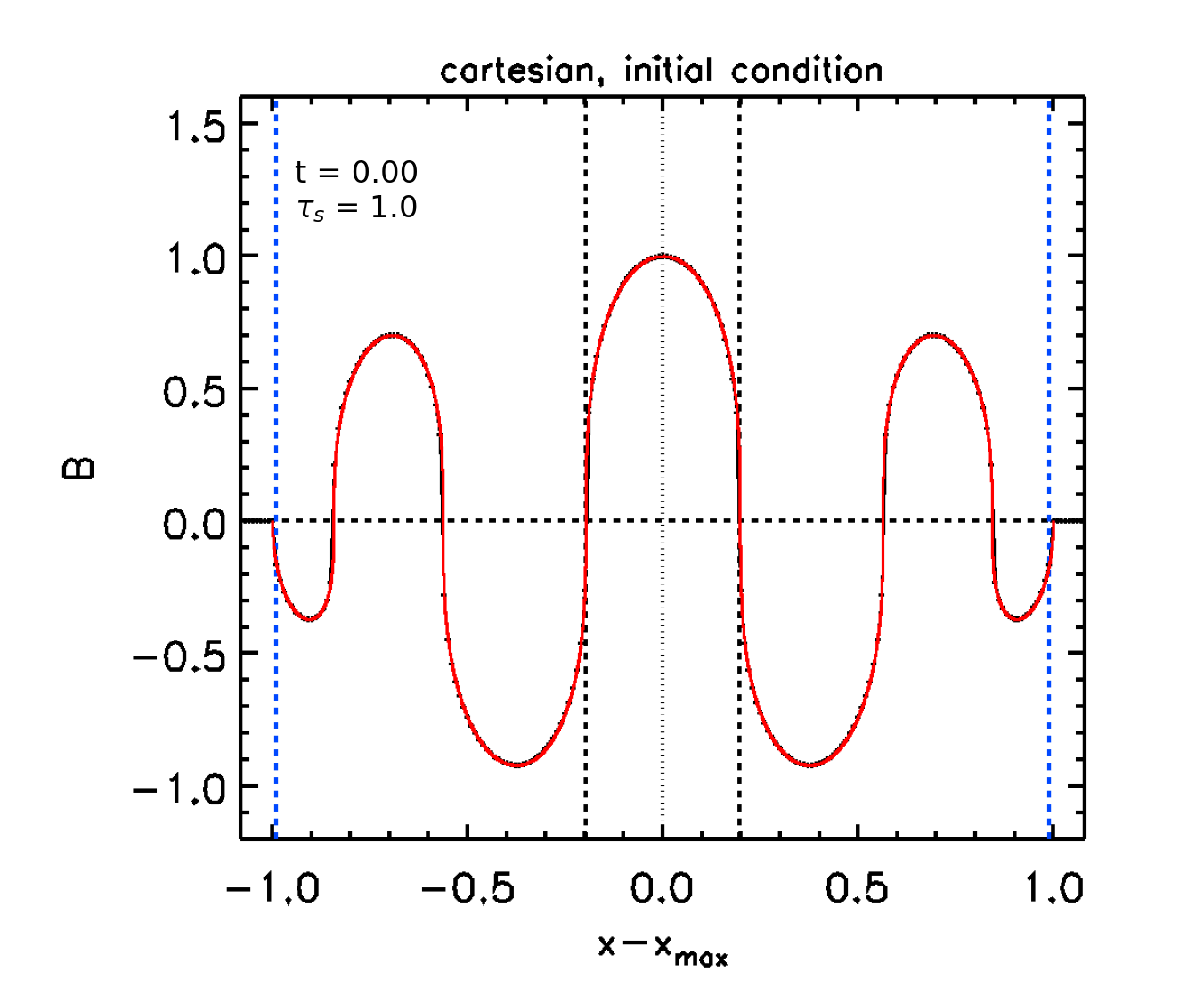}\hskip -6mm
\includegraphics[width=0.38\textwidth]{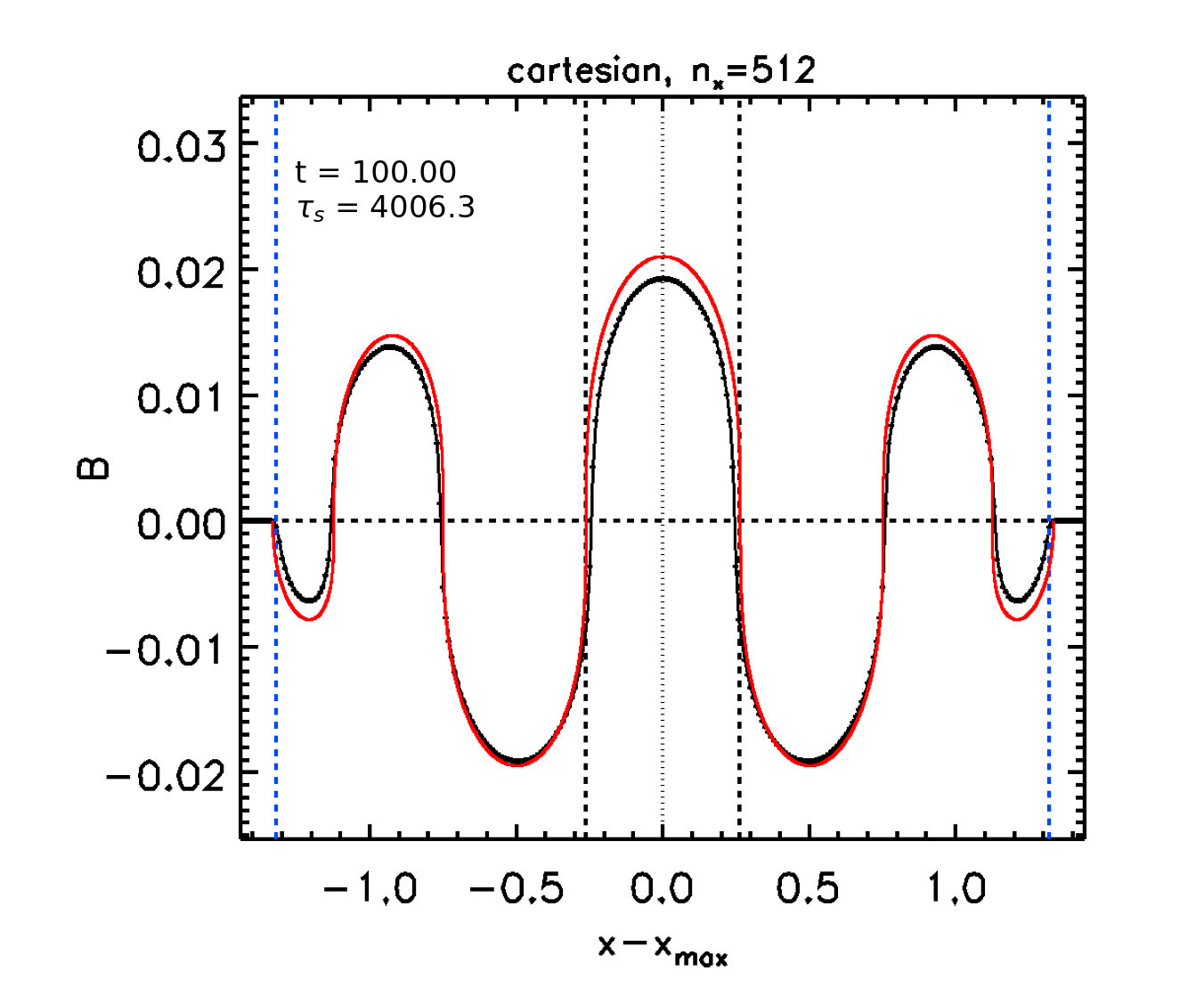}\hskip -6mm
\includegraphics[width=0.37\textwidth]{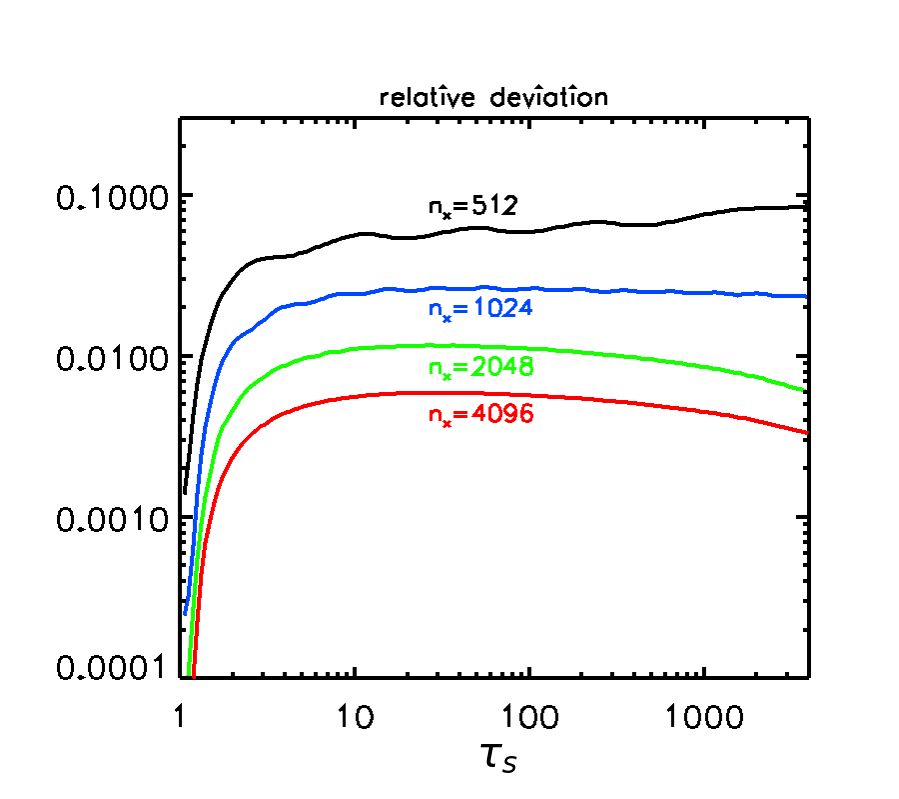}}
\caption{A comparison of the Bifrost solution (black thick curve) for the third
  harmonic with the theoretical solution (red curve), showing  the initial  (left) and  final (middle) profiles for a resolution of  $n_x=512$. Right panel: the relative mutual deviation
  of the maxima between the theoretical and Bifrost solutions for the four
  resolution levels used in the test.}
\label{fig:third_harmonic_main}
\end{figure*}

For this test we have used four levels for the spatial resolution,
corresponding to $n_x=512, 1024, 2048$ and $4096$ grid cells across the
domain. The grid spacing is $\Delta x = 0.01$ for the worst resolution and
$\Delta x = 0.00125$ for the best one. In Fig.~\ref{fig:bifrost_cartesian},
the initial function for all the cases is shown in the left panel; the final
state at time $t=100$ ($\tauselfsimilarexpansion = 1206$) for the lowest-resolution case
($n_x=512$) is presented in the middle panel.  
The theoretical solution for the first harmonic, extended with
its mirror-symmetric solution for negative $\xi$, is superimposed as a thick
red curve. By construction, at time $t=0$ (left panel) the red and black curves lie exactly on top of
each other. In the middle panel, we notice a small
mutual deviation at the final time.  The absolute value of the relative
deviation between the maxima of the theoretical and the Bifrost solutions for
each of the resolution levels is shown in the right panel as a function
of time.  The relative deviation first grows and then stays at a roughly
constant level in the advanced stages: this is due to the improvement of
spatial resolution that naturally comes about through the expansion of the
spatial support of the function as time progresses.

Another way of testing the Bifrost solution is to compare with the power laws  (\ref{eq:B_R_constraint}) -- (\ref{eq:tau_s}) for the decay of $\myBzero$ and expansion of $\myL$ given by  $K$ and
$H$. Thus, for the first harmonic from Table~\ref{tab:one}, the exponents of
the power laws are $\alpha = -0.387$ and $\gamma = 0.113$. 
Fig.~\ref{fig:power_laws_cartesian} gives a comparison for the worst-resolution case $n_x=512$ (black solid: Bifrost solution; red dashed: theoretical laws). The left panel gives the  comparison for $\myBzero(t)$, while the position of the outer front, $\myL(t)$, and of the internal current sheet, $\myL_{cs}(t)$, are given in the central panel. The a-posteriori determination of the location of the outer front is prone to numerical inaccuracies, due to the sharp corner between the solution (which declines like $|\,2\,H\,\delta\,|^{1/2}$, see Section~\ref{sec:symmetric}) and the zeros beyond the front. On the other hand, the time-dependent calculation of the location of the current sheets is most demanding for the Bifrost runs, as explained in the preamble to this section. 
Yet, the figure shows the match between theory and Bifrost results  to be excellent: the Bifrost curves agree with those of the theoretical law
(\ref{eq:B_R_constraint}) -- (\ref{eq:tau_s}) with a precision of
$\pot{5}{-3}$ for $\myBzero(t)$ (left panel), $\pot{2}{-2}$
(middle panel, 
$\myL$) and $10^{-2}$ (middle panel, $\myL_{cs}$). 
 The highest resolution case ($n_x=4096$) gives a precision of
 $\pot{7}{-4}$ for 
${\myBzero}(t)$, $\pot{3}{-3}$ for $\myL(t)$, and $\pot{2}{-3}$ for
$\myL_{cs}(t)$, respectively. 

The ability of Bifrost to cope with the singularities at the internal
  current sheets can be further tested by evaluating the integral of the magnetic field
in either of the lobes and studying how it varies in time.  The integral of
the magnetic field between the nulls or between the central maximum and the current sheet
must evolve in time exactly as $\myBzero(t)\,\myL(t)$, i.e., using the analytical
laws~(\ref{eq:tau_s}), as a power-law of $\tauselfsimilarexpansion(t)$ with exponent
$(1-\Ktilde)/[2(1+\Ktilde)]$. Looking at the value of $\Ktilde$ in
Table~\ref{tab:one} for the different harmonics, we see that the 
integral is conserved in time for the ZKBP solution ($\Ktilde = 1$) and
decreases for any of the other harmonics. For the first harmonic, in
particular, $(1-\Ktilde)/[2(1+\Ktilde)] = -0.275$. The comparison between
numerical solution and analytical power law appears in the right panel of
Fig.~\ref{fig:power_laws_cartesian}, again for the worst-resolution case
($n_x=512$). The power-law exponent of the minimum-square fit to the
numerical curve matches the analytical value with a precision
$\pot{7}{-3}$. For the $n_x=4096$ case, the precision is an order of
magnitude better, $\pot{9}{-4}$.

As a final remark, we note that the harmonic used in this section is expected
to be stable (as discussed in Sec.~\ref{sec:stability_1st_eigenmode}). This
implies that the solution of the equation, even if with the addition of small
zero net-flux perturbations, naturally, i.e., mathematically, tends to the
exact harmonic. Still, if the numerical integration at the sharp current
sheets were inaccurate, then magnetic flux would be artificially annihilated
across them, and the timescale of evolution could deviate importantly from
that given by the power laws of
Eq.~(\ref{eq:B_R_constraint})-(\ref{eq:exponents}). Also, depending on the
integrator used, the numerical errors could cause an artificial deviation of
the net flux from zero. The fact that the fit to the theoretical solution is
very good is the best proof of the accuracy of the Bifrost calculation.

\subsection{Higher harmonics}\label{sec:bifrost.2}

Consider the third symmetric harmonic (light-green curve in
Fig.~\ref{fig:the_cartesian_solutions}), which contains six internal nulls
(three on each side of the maximum) as well as the two outer fronts.  The
properties of the theoretical solution (Sec.~\ref{sec:self-similar},
table~\ref{tab:one}) suggest that the decay of the maxima with
$\alpha=-0.465$ is faster than for the fundamental, whereas the spatial
expansion is much slower with $\gamma = \pot{3.47}{-2}$.  

The Bifrost solution for the lower-resolution case ($n_x=512$) is shown in
Fig.~\ref{fig:third_harmonic_main}, where it can be observed that the
  deviation from the theoretical solution is more pronounced than for the
first symmetric harmonic (Fig.~\ref{fig:bifrost_cartesian}). The effective
resolution for each lobe is roughly twice as bad as for the first
  harmonic.  The same is true for the corresponding curves for the maxima
(right panel of the figure).  For the highest-resolution case ($n_x = 4096$),
the maximum relative deviation is quite small ($\pot{6}{-3}$) but
is a factor of $3$ larger than for the first harmonic. The low-resolution
case gives a relative deviation of several percent, and continuously
deteriorates in the right half of the $\log\tauselfsimilarexpansion$ range.

  The location of the zeros and of the corresponding power-law exponents
  provides a challenging test in the lower-resolution case ($n_x=512$) for
  the high harmonics. The numerical location of the outer front 
[$\myL(t)$] is found to have a maximum deviation of $2\%$ from the theoretical value,
  while the fit of $\myL(t)$ to a power law agrees with the theoretical value
  to $6\%$.  The situation is more demanding for the innermost current sheet,
  since, although $\myL_{cs}^{}(t)$ is determined numerically with accuracy
  better than $10\%$, its fit to a power law produces an exponent that agrees
  with the theoretical value only within $\pot{2}{-1}$.  However, the
  maximum of the solution and the magnetic flux in the innermost lobe yield
  exponents with an acceptable accuracy of $\pot{2}{-2}$ of the theoretical
  values.  Naturally, the higher-resolution cases produce better comparisons
  between theoretical and numerical variables.  The highest-resolution case
  ($n_x=4096$) gives an accuracy of $\pot{3}{-4}$ for the power-law
  exponent of the decay of the maximum, $10^{-2}$ for the location of the
  outer front, $10^{-2}$ for the location of the innermost current sheet, and
  $\pot{3}{-4}$ for the integrated flux. 

To properly judge the value of the test for this harmonic or higher-order
ones, one has to take into account that it is unstable (like all harmonics
above the first) , i.e., prone to decay into the first symmetric harmonic (or
the dipole solution, if antisymmetric cases are considered) whenever any zero
net-flux deviation is added to it, as can be easily checked using a simple
integrator, like that used in Sec.~\ref{sec:self-similar}. So, the question
is: when is that transformation going to take place? If the accuracy of the
initial conditions and of the numerical integration were extremely high, one
would expect the transformation to occur for very large diffusive
times. Indeed, using the integrator of Sec.~\ref{sec:self-similar}, one sees
that the first loss of nulls from the solution (either through merging of
pairs of nulls or ejection through the outer boundary) occurs at times
$t=\pot{2.4}{7}$, $t=\pot{6.6}{7}$, and $t=\pot{5.5}{9}$ ($\log\tau_s = 9.0,
9.4, 11.3$) for resolutions of, respectively, $1024$, $2048$ and $4096$ cells
in the whole domain of integration. So, in a certain sense, this test is more
demanding than that for stable harmonics: additionally to the problem of the
possible artificial flux annihilation at internal current sheets (see
Sec.~\ref{sec:bifrost.1}), which can modify the time- or length-scales of the
solution while preserving the shape of the stable harmonic, the solution here
is also intrinsically prone to transformation towards a different harmonic.
By the same token, the test in this case consists basically in checking for
how long the numerical solution remains in the close neighbourhood of the
initial harmonic without any apparent transformation toward the stable
lower-order harmonic.

\section{Discussion}
\label{sec:discussion}

Increasingly, the importance of ambipolar diffusion in the solar atmosphere
is being realised and incorporated in advanced multidimensional
codes  such as 
Bifrost \citep{martinez12a,Nobrega-Siverio_etal_2020a}, 
MURaM \citep{Cheung_Cameron:2012,Rempel_Przybylski:2021}, 
MANCHA \citep{khomenko12a,Gonzalez-Morales_2018}, 
MPI-AMRVAC \citep{Popescu_Keppens_2021,Keppens_etal:2023}, 
Lare3D \citep{Chouliaras_etal:2023}, 
NIRVANA \citep{Ziegler:2011}, and others.
The fact that ambipolar diffusion is a nonlinear process leads to much
unexpected behaviour that differs from the familiar
linear process of Ohmic diffusion.  Understanding the physical
effects of ambipolar diffusion is therefore of importance, and one way to
disclose them is through a study of the pure diffusion
problem, as carried out in this paper for the Cartesian case and in our
preceding paper \citep{Moreno-Insertis_etal_2022} for the 2D cylindrical
case.

\subsection{Simple 1D solutions}

The importance of singular profiles of $B(x) \propto x^{1/3}$ type has been
discussed. For static situations, a central Ohmic-diffusion region resolves
the singularity of the field profile near the null point; however, an
important realisation by \citet{Parker_1963} is that the overall diffusion
timescale, the magnetic flux transfer rate toward the null, and also the rate
of Ohmic heating are determined not by Ohmic diffusion near the null but by
the ambipolar diffusion profile.  On the other hand, we have found
stagnation-point flow solutions in which an external, pure advection region
surrounds an internal ambipolar diffusion region, and the latter contains in
its core a small domain dominated by Ohmic diffusivity.  Oppositely directed
magnetic flux is transported in from large distances first through pure field
advection until it reaches the diffusive central region. This consists of:
first, an ambipolar-diffusion dominated domain, with the field steepening to
a singular $\propto x^{1/3}$ profile; then an Ohmic diffusion region, in the
immediate surroundings of the central null, where the field is
annihilated. The flux transfer rate is uniform throughout the entire
  domain and, as in the classical stagnation flow problem, it is determined by
  the value of the electric field.

\subsection{ Microscopic aspects}

Concerning microscopic (i.e., multi-fluid) aspects, a word of caution
  is necessary when dealing with singular profiles created by ambipolar
  diffusion effects, especially if the singularities are tied to fixed plasma
  elements like those of Sec.~\ref{sec:general_solutions}
or the central singularity in the antisymmetric eigenmodes of
Sec.~\ref{sec:antisymmetric}.  
As discussed by \citet{Zweibel_1994} and \citet{Brandenburg_Zweibel_1994} for
the $B(x) \propto x^{1/3}$ profiles, 
in those cases, one has to distinguish the bulk plasma speed from the motion
of its individual species. In the standard ambipolar-diffusion approach
(see~Sec.~\ref{sec:introduction}), the effect of the Lorentz force on the
ions, but not on the neutrals, leads to a mutual drift and to a drift of
either of them with respect to the bulk plasma motion; in a 1D situation with
$\Bvec=B(x,t)\, \uvec{y}$, the latter can be written 
\begin{equation}\label{eq:drift_speed_1D}
    v_i - v = -\,\mu_0\,\chi_a  \,
\fracdps{\partial }{\partial x}\left(\fracdps{B^2}{2\, {\mu_0^{}}}\right)
\hbox{\hskip 3mm and\hskip 3mm}
  v_n -v = - \;\fracdps{\rho_i}{\rho_n}\; (v_i - v)\;.
\end{equation}
In a $B \propto x^{1/3}$ profile with zero bulk plasma speed, even
if resolved by an Ohmic layer near the null, these drift speeds cause the
accumulation of ions (and also the depletion of neutrals) around the null,
nearing a singular behaviour if the Ohmic layer is tiny by comparison to the
ambipolar-dominated region. This can be a problem when the profile is fixed
at the same plasma elements.
In standard kinematic models 
all dynamical considerations are set aside, and, coherently, problems related
to accumulation of microscopic particles could also be ignored;
nevertheless, this neglect becomes particularly critical in cases when the
ambipolar-diffusion dominated range reaches very close to the
singularity. So, while the solutions in this paper and others are of interest
when the microscopic aspects can be ignored because the time
evolution of the system under consideration renders them unimportant (and,
also, as mathematical solutions and as demanding tests for MHD codes), the
multi-fluid aspects should be kept in mind in critical situations when
near-singularities are reached and maintained at the same plasma
  elements for timescales comparable to other physical timescales of the
problem.

\subsection{Self-similar solutions}

Self-similar solutions have been studied here in detail, with magnetic fields
$\Bz(x,t)=g(t)f(\xi)$ that depend on the spatial coordinate in the
combination $\xi=x/\myL(t)$; they are appropriate when there are no natural
length or timescales in the system, so all that can be expected is that the
latter fulfill mutual relations like those of
Eqs.~(\ref{eq:B_R_constraint})--(\ref{eq:exponents}).  The zero-order
solution and the first symmetric and antisymmetric harmonics are the basic
solutions of the problem.  The zero-order solution is the only one with
non-zero net flux; the first, or higher, harmonics have as many internal null
points in the $x \ge 0$ domain as the order of the harmonic. At those
internal nulls, positive magnetic flux is brought in toward the null from one
side and negative flux from the other, so that flux annihilation ensues.  In
this way, while the net zero flux of the eigenmode is preserved in the time
evolution, the unsigned flux gradually decreases in time toward zero. This
annihilation follows from the cancellation of the exponents of $\Bz^2$ and
$d\Bz/dx$ in the expression for the flux transfer rate
(Eq.~\ref{eq:diff_flux_definition}, with $\eta=0$) for $\Bz\propto x^{1/3}$
and is of interest in the absence of any Ohmic diffusion, i.e., from a purely
mathematical point of view.  In fact, with a small, but non-zero Ohmic
dissipation, the annihilation will be Ohmic and take place in the immediate
neighbourhood of the null points. Still, the solution
outside that neighbourhood will be the exact harmonic of the pure ambipolar
diffusion problem.

\subsection{Mode stability}

We have probed the stability of a few of the eigenmodes through calculations
starting with an unperturbed high-order eigenmode, or with a perturbed
low-order one. When the net flux of the initial condition is zero, we have
seen how the solution evolves toward either the first symmetric harmonic or
the dipole solution. Simple cases with non-zero initial net flux lead to the
fundamental ZKBP solution with the same net flux: this elementary case has
not been illustrated here but has been repeatedly obtained by us in different
calculations and is well known from the literature. The cases with zero
initial net flux are more interesting. The dipole mode is an attractor; the
first symmetric mode is likely to be one but, to our knowledge, this has not
yet been fully proved with mathematical tools. What we have obtained here is
just an {\it empirical} hint, so to say, that the higher-order eigenmodes
tend to evolve into the first-order harmonics (either the symmetric or
antisymmetric one) via cancellation of the internal null points in pairs or
ejection of single nulls through the outer boundary; we have also shown, in
one of the examples, that an asymmetric perturbation of the first symmetric
harmonic can make it turn into the dipole solution asymptotically in time.

\subsection{Bifrost and large-code tests}

We are proposing that the time evolution of the eigenmodes of the
self-similar problem can be used as strong tests for ambipolar diffusion
solvers in MHD codes. Given the probably unstable nature of the higher-order
modes, two types of simple test can be envisaged: either a pure first-order
harmonic is verified to maintain its shape in time according to the
self-similar laws, or, second, a higher-order harmonic can be seen to evolve
into the first-order eigenmodes asymptotically in time. We have carried out
both types of test with the Bifrost code.  For the second one
(Sec.~\ref{sec:bifrost.2}), which is less amenable to quantitative
evaluation, we have checked the time evolution of the third symmetric
harmonic. Here, for a comparatively low-resolution case ($n_x=512$ cells),
the code has kept the 3rd-harmonic appearance at least until $t=100$ ($\log
\tau_s = 3.6$), with roughly twenty times higher accuracy for $n_x=4096$
cells. For the first type of test (Sec.~\ref{sec:bifrost.1}), we have used
the first symmetric harmonic as initial condition; in the Bifrost run, the
solution maintained its self-similar appearance in time to a high order of
accuracy, both concerning the deviation from the exact mathematical solution
for it as well as the integral of the flux in its central lobe and the
location of the internal nulls. It is worth noting the difficulty of passing
this test for the code: in the exact solution, the internal nulls are an
essential singularity; in the numerical solution, that singularity will be
resolved by ohmic diffusivity (or by the intrinsic numerical diffusivity of
the code). The fact that the null and its associated singularity evolve in
time in the Bifrost calculation as prescribed by the theoretical solution can
be taken as an indicator of the quality of the ambipolar diffusion solver in
the code.

\section{Conclusions}\label{sec:conclusions}

\begin{itemize}

\item Solutions have been considered in which an ambipolar diffusion
  region harbours in its core an Ohmic-diffusion one; in particular, a
  generalisation of the classical stagnation-point flow has been found with
  three domains (advection-, ambipolar- and Ohmic-dominated), where each is
  contained in the deep interior of the previous one.  Flux annihilation
  takes place in the Ohmic region, but at a rate determined by the advection
  flow in the external domain and mediated by the ambipolar diffusion region.

\item We have calculated the parameters and runs of the eigenmodes for the
  self-similar solutions of the 1D Cartesian ambipolar diffusion equation,
  i.e., of the countable subset of self-similar solutions which have compact
  support. These solutions have been efficiently obtained using phase-plane
  techniques in spite of the essential singularities they have within their
  domain. Although their existence had previously been studied in the
  mathematical literature, their form had not yet been presented.

\item The stability of those eigenmodes has been probed using direct
  integration of the time-dependent 1D ambipolar diffusion equations. In
  particular, we have seen that modes with order above the first, or modes
  with zero net-flux perturbations, spontaneously evolve into the
  first-order modes. Although not valid as a stability proof in the
  mathematical sense, those results can be used as a strong suggestion (and
  visualization) of the unstable nature of the higher-order modes. On the
  other hand, the behaviour of the first-order symmetric and antisymmetric
  eigenmodes in the numerical solutions is consistent with the expectation
  that they are stable.

\item 
Tests for the ambipolar diffusion solver in use in the Bifrost code
  have been carried out, one with the first symmetric harmonic, and another
  with a higher-order symmetric mode that is expected to transform
  spontaneously in the first one. The tests were successful. Beyond that, the
  usefulness of such tests for other MHD codes has been discussed.
\end{itemize}

\begin{acknowledgements}
This research has been supported by the European Research Council through the
Synergy Grant number 810218 (``The Whole Sun'', ERC-2018-SyG); by the Spanish
Ministry of Science, Innovation and Universities through project
PGC2018-095832-B-I00; and by the Research Council of Norway (RCN) through its
Centres of Excellence scheme, project number 262622. Use of the resources provided by Sigma2 - the National Infrastructure for High Performance Computing and Data Storage in Norway is also acknowledged. 
\end{acknowledgements}

\bibliographystyle{aa}

\end{document}